\documentclass[10pt,twocolumn]{article}

\usepackage[utf8]{inputenc}
\usepackage[T1]{fontenc}
\usepackage{times}
\usepackage[top=0.75in,bottom=0.75in,left=0.6in,right=0.6in]{geometry}
\usepackage{graphicx}
\usepackage{amsmath,amssymb}
\usepackage[colorlinks=true,linkcolor=blue,citecolor=blue,urlcolor=blue]{hyperref}
\usepackage{titlesec}
\usepackage{fancyhdr}
\usepackage{caption}
\usepackage{enumitem}
\usepackage{dblfloatfix}
\usepackage{xcolor}
\usepackage{bm}
\usepackage{tabularx}
\usepackage[section]{placeins}  

\titleformat{\section}{\normalsize\bfseries}{\thesection}{0em}{}
\titleformat{\subsection}{\normalsize\bfseries\itshape}{\thesubsection}{0em}{}
\titlespacing{\section}{0pt}{10pt}{4pt}
\titlespacing{\subsection}{0pt}{8pt}{2pt}

\let\OLDthebibliography\thebibliography
\renewcommand\thebibliography[1]{%
  \OLDthebibliography{#1}%
  \setlength{\parskip}{0pt}%
  \setlength{\itemsep}{0pt plus 0.3ex}%
}

\begin{document}

\twocolumn[{%
\begin{center}
  {\LARGE\bfseries Computing with the complex nonlinear dynamics of an optomechanical oscillator\par}
  \vspace{10pt}
  {\normalsize
  Shulamit~Edelstein$^{1,\dagger}$*,
  Marcos~Men\'endez$^{1,\dagger}$,
  Bingrui~Lu$^{2}$,
  Babak~Vosoughi~Lahijani$^{2}$,\\[2pt]
  Cefe~L\'opez$^{1}$,
  Miguel~C.~Soriano$^{4}$,
  S\o{}ren~Stobbe$^{2,3}$,
  Pedro~David~Garc\'ia$^{1}$*\par}
  \vspace{6pt}
  {\small\itshape
  $^{1}$Instituto de Ciencia de Materiales de Madrid (ICMM-CSIC),
  Sor Juana In\'es de la Cruz 3, 28049 Madrid, Spain\\
  $^{2}$Department of Electrical and Photonics Engineering,
  Technical University of Denmark, DK-2800 Kgs.\ Lyngby, Denmark\\
  $^{3}$NanoPhoton -- Center for Nanophotonics,
  Technical University of Denmark, DK-2800 Kgs.\ Lyngby, Denmark\\
  $^{4}$Instituto de Física Interdisciplinar y Sistemas Complejos (IFISC), CSIC--UIB,
Campus UIB, 07122 Palma de Mallorca, Spain\\[2pt]
  *Corresponding authors: shulamit.edelstein@csic.es; pd.garcia@csic.es\\
  $^\dagger$These authors contributed equally to this work.}
  \vspace{10pt}

  \parbox{0.92\textwidth}{\small\bfseries
An optomechanical oscillator undergoes a Hopf bifurcation that connects two dynamical regimes with different information-processing capabilities: thermal Brownian motion and coherent self-sustained oscillation. Below threshold, the oscillator occupies a stable fixed point around which thermal fluctuations drive stochastic Brownian motion — a regime dominated by linear response, with only short-lived memory and negligible usable nonlinearity. Above threshold, radiation pressure, free-carrier dynamics, and thermo-optic relaxation act together to sustain a stable limit cycle that simultaneously provides both nonlinear transformation and dynamical memory. Here we show that this coherent regime can be used as a physical reservoir for computation: by perturbing the phonon-lasing attractor, the cavity performs nonlinear input--output transformations and retains short-term memory without any external feedback mechanism. Using only a single chip-integrated device with 20 virtual nodes, we reconstruct nonlinear functions, predict the evolution of chaotic time series, and perform spoken digit classification on a two-digit sub-task. The mechanical resonance frequency sets the intrinsic dynamical timescale of the reservoir and therefore its processing speed; while the present device operates near 0.4~GHz, optomechanical and nanomechanical systems can be engineered to reach multi-GHz and sub-terahertz frequencies, directly translating into a scalable path toward ultrafast integrated physical computing.}

  \vspace{8pt}
\end{center}
}]

\begin{figure*}[t]
\centering
\includegraphics[width=0.9\textwidth]{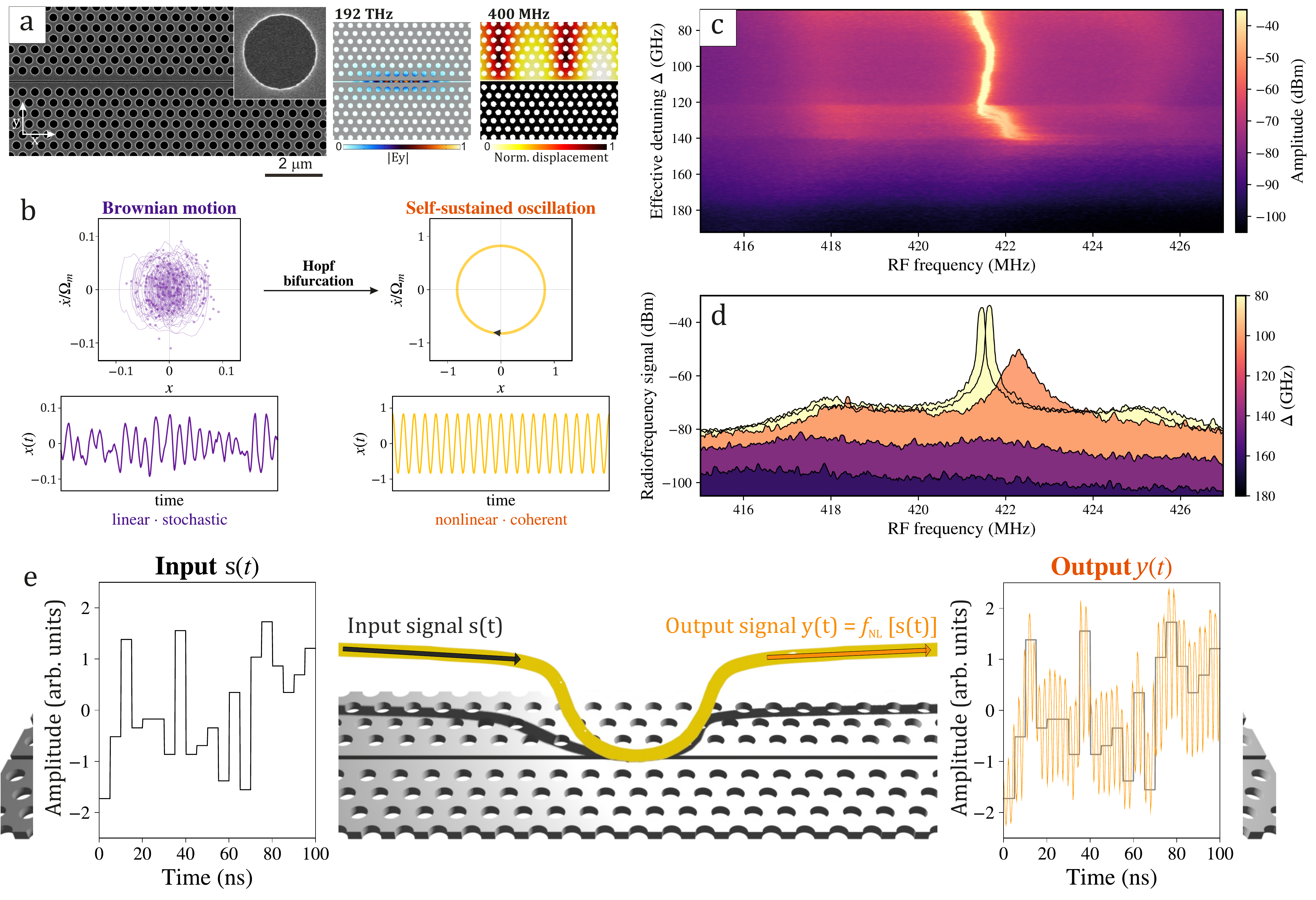}
\caption{%
\textbf{An optomechanical oscillator through a Hopf bifurcation as a physical reservoir.}
(\textbf{a})~Scanning electron micrograph of the silicon photonic-crystal slot-waveguide cavity. Overlaid: optical mode profile ($|E_y|$) and normalized mechanical displacement. (\textbf{b})~Two dynamical regimes. Left: Brownian motion---noisy phase portrait and stochastic time trace (linear, stochastic). Right: self-sustained oscillation---limit cycle and coherent time trace (nonlinear, coherent). Arrow indicates the Hopf bifurcation. (\textbf{c})~Radiofrequency power spectrum as a function of the effective laser--cavity detuning $\Delta$, showing the transition from broad thermal noise to a sharp coherent peak. (\textbf{d})~Line cuts at selected detunings. (\textbf{e})~Reservoir computing scheme: input signal $s(t)$ modulates the optomechanical cavity; the cavity output $y(t)$ is read out via a slot waveguide evanescently coupled to a tapered fiber.}
\label{fig:1}
\end{figure*}

\begin{figure*}[t]
\includegraphics[width=1\linewidth]{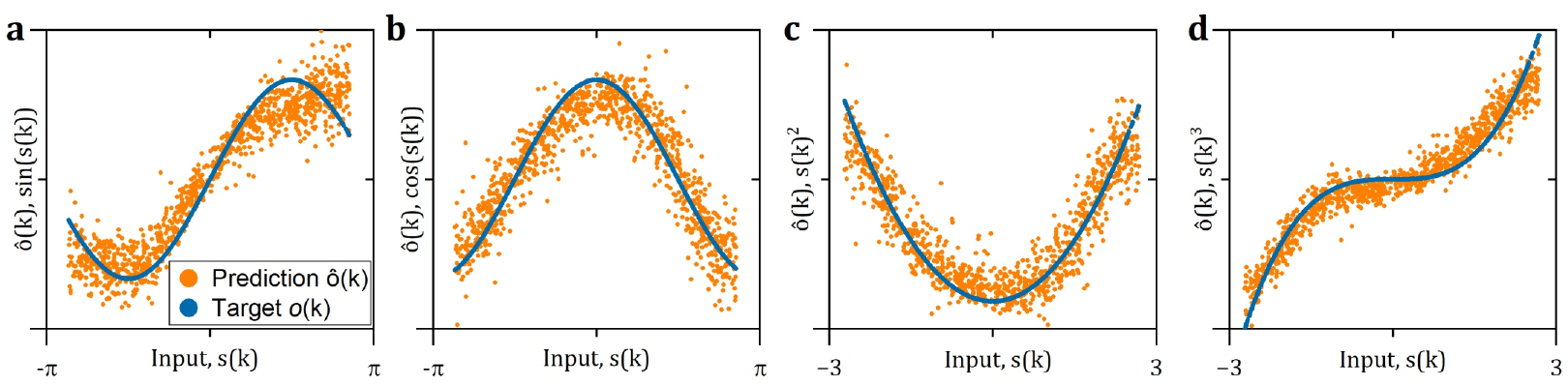}
\caption{%
\textbf{Nonlinear transformation provided by the optomechanical reservoir.}
Reconstruction of four nonlinear target functions using 20 virtual nodes:
(\textbf{a})~$\sin(s(k))$,
(\textbf{b})~$\cos(s(k))$,
(\textbf{c})~$s(k)^2$, and
(\textbf{d})~$s(k)^3$.
The input $s(k)$ consists of uniformly distributed random values.
Dots represent the reservoir prediction $\hat{o}(k)$ obtained from a linear readout trained on the time-multiplexed cavity states; solid lines show the exact target functions.
The accurate reconstruction shows the intrinsic nonlinear mapping of the optomechanical dynamics.
The odd target functions ($\sin$, $s^3$) are systematically better reconstructed than even ones ($\cos$, $s^2$); the physical origin of this asymmetry under power modulation is discussed in Supplementary Section~\ref{sec:NL}.}
\label{fig:2}
\end{figure*}

A Hopf bifurcation in an optomechanical oscillator~\cite{Aspelmeyer2014} sets a boundary between two dynamical regimes with different information-processing capabilities. Below threshold, the mechanical oscillator occupies a stable fixed point around which thermal fluctuations drive stochastic Brownian motion. In this regime, the response is dominated by linear dynamics and noise, while any nonlinear transduction arising from optomechanical backaction remains weak and incoherent. This is the regime of molecular motors and DNA replication that Bennett showed can support thermodynamically reversible computation~\cite{Bennett1982}, illustrating that information processing can occur even in stochastic systems. Above threshold, radiation pressure, free-carrier dynamics, and thermo-optic relaxation act together to destabilize the fixed point and sustain a coherent limit cycle. The resulting dynamics are nonlinear and coherent, and the oscillator acquires finite dynamical memory: the mechanical response to a perturbation persists over multiple oscillation cycles. We exploit these three properties---nonlinearity, coherence, and memory---for reservoir computing~\cite{Jaeger2004,Maass2002}.

\begin{figure*}[t]
\centering
\includegraphics[width=1\textwidth]{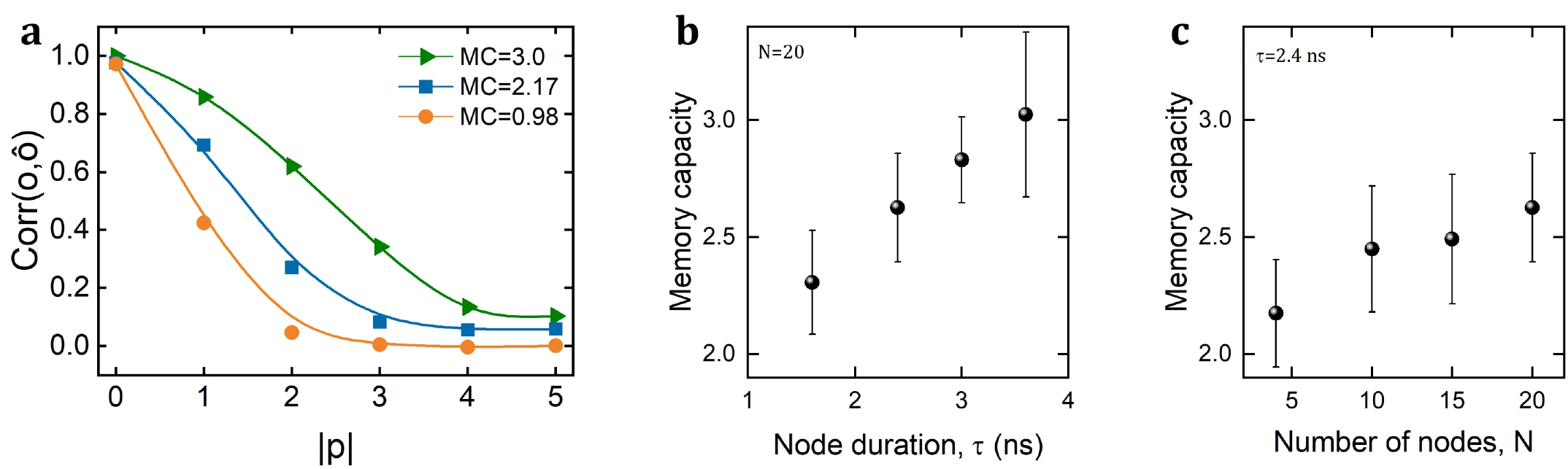}
\caption{%
\textbf{Experimental memory capacity of the optomechanical reservoir.}
(\textbf{a})~Fading memory diagrams showing the Pearson correlation between past input values and their reconstruction by the reservoir, for three representative values of the linear memory capacity: MC~$\approx 0.98$ and MC~$\approx 2$ ($N=4$, $\tau=2.4$~ns), and MC~$\approx 3$ ($N=20$, $\tau=3.6$~ns). The correlation decays with the number of past steps $|p|$; reservoirs with higher memory capacity retain significant correlations over more steps. Solid curves are guides to the eye.
(\textbf{b})~Memory capacity as a function of node duration time $\tau$ at fixed number of nodes $N=20$. Increasing $\tau$ allows the coherent oscillation to integrate information over more mechanical cycles, systematically increasing memory. Error bars indicate the standard deviation over repeated measurements.
(\textbf{c})~Memory capacity as a function of the number of virtual nodes $N$ at fixed $\tau=2.4$~ns. Performance converges with modest reservoir size, consistent with the linear memory capacity being more limiting than reservoir dimensionality in the explored range. Error bars indicate the standard deviation over repeated measurements. Numerical simulations of the memory capacity as a function of mechanical frequency and node duration are presented in Supplementary Fig.~S11.}
\label{fig:3}
\end{figure*}

Reservoir computing is a computational framework, often realized as a recurrent neural network, in which a fixed nonlinear dynamical system---the reservoir---maps input signals into a high-dimensional state space from which a simple linear readout is trained to approximate target functions~\cite{Lukosevicius2009,Tanaka2019}. Because the reservoir connections are fixed, the approach lends itself to direct implementation in physical systems whose intrinsic nonlinear response can be exploited at no additional energy cost. Photonic reservoirs have been realized in spatially distributed networks of waveguides and splitters~\cite{Vandoorne2014}, parallel arrays of semiconductor lasers~\cite{heuser2020}, and time-delay architectures based on fiber loops~\cite{Brunner2013,Larger2017,Appeltant2011}, reaching processing speeds above 60~GHz in silicon photonic chips~\cite{Wang2024,Yan2024}. In all of these implementations, however, nonlinearity and memory reside in separate physical subsystems---nonlinearity in the gain medium or at detection, memory in an external feedback path~\cite{abdalla2026}---which adds energy overhead, dominates the device footprint, caps the processing rate at the feedback timescale rather than the intrinsic nonlinear one, and prevents joint engineering of the three properties since they are controlled by independent parameters. The optomechanical reservoir presented here removes this separation: nonlinearity, memory, and dynamical timescale all emerge from the same physical interaction within a single chip-integrated cavity, with no external feedback. The mechanical resonance frequency---set by the slab geometry at fabrication---then provides a single physical parameter that simultaneously sets all three.

\begin{figure*}[t]
\centering
\includegraphics[width=1\textwidth]{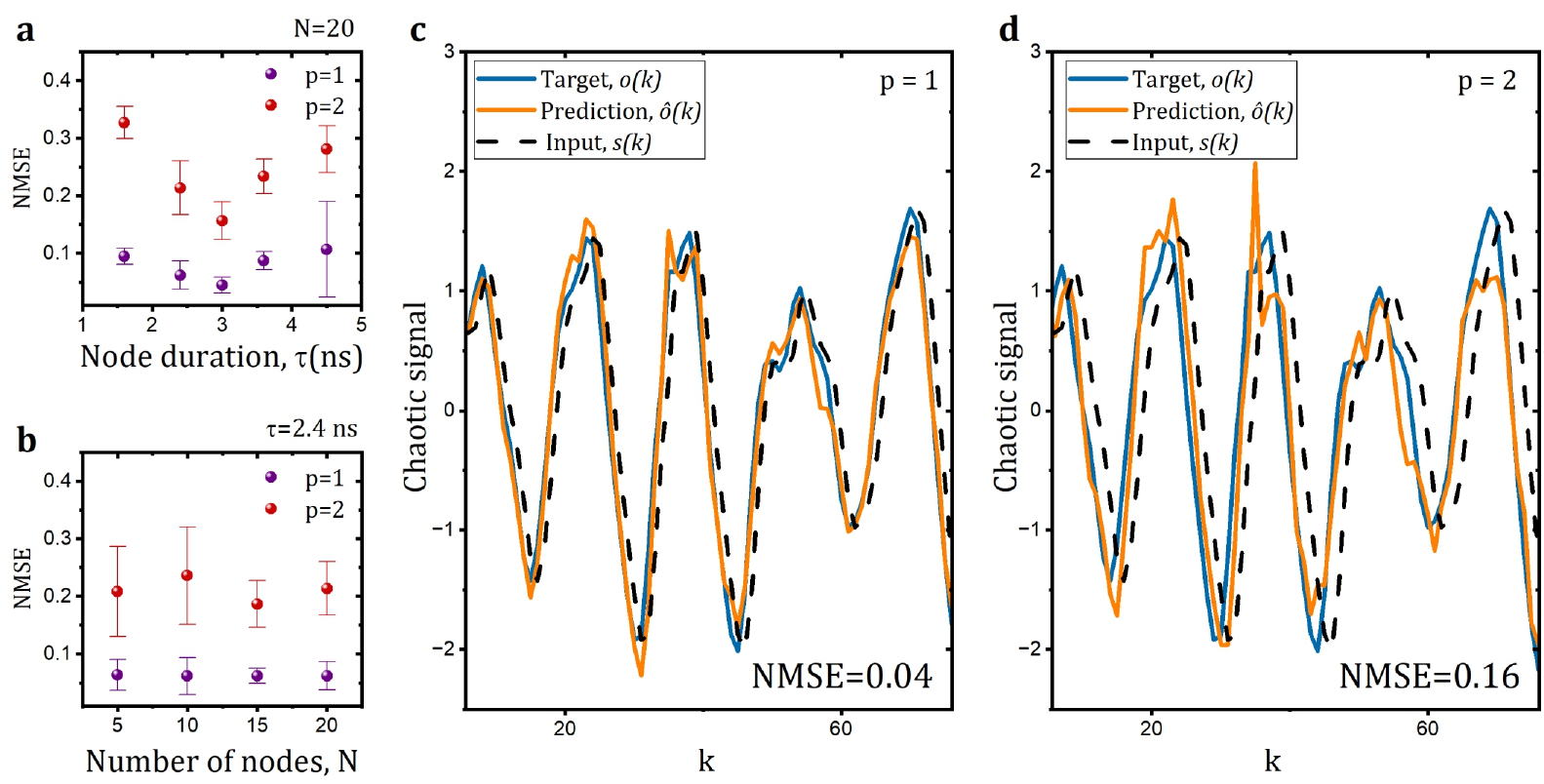}
\caption{%
\textbf{Prediction of the Mackey--Glass chaotic time series.}
(\textbf{a})~Normalized mean square error (NMSE) as a function of node duration $\tau$ for one-step ($p=1$, purple) and two-step ($p=2$, red) prediction using 20 virtual nodes. The error first decreases with increasing $\tau$ and reaches a minimum at $\tau = 3$~ns for both prediction horizons, corresponding to the optimal memory depth within the explored range, as the cavity state integrates information over multiple mechanical cycles. The persistent separation between $p=1$ and $p=2$ across all node durations reflects the finite predictability horizon imposed by mechanical damping and thermal dissipation. Error bars indicate the standard deviation over repeated measurements.
(\textbf{b})~NMSE as a function of the number of virtual nodes $N$ at fixed $\tau = 2.4$~ns, for one-step ($p=1$, purple) and two-step ($p=2$, red) prediction. Performance converges with modest reservoir size and saturates beyond $N \approx 10$--15, confirming that the limiting factor is the linear memory capacity rather than the dimensionality of the reservoir embedding. Error bars indicate the standard deviation over repeated measurements.
(\textbf{c},~\textbf{d})~Representative chaotic time traces for $p=1$ and $p=2$, respectively: input $s(k)$ (black dashed), target $o(k)=s(k+p)$ (blue), and reservoir prediction $\hat{o}(k)$ (orange). The reservoir reproduces both amplitude and phase for short prediction horizons, while deviations accumulate for larger $p$, consistent with the finite fading memory set by the mechanical and thermal timescales. All computations use only the intrinsic nonlinearity and dynamical memory of the optomechanical cavity, with no external feedback.}
\label{fig:4}
\end{figure*}

Silicon photonic-crystal slot-waveguide cavities simultaneously confine telecom-wavelength light and support mechanical extended modes near 0.4~GHz at room temperature~\cite{Arregui2023} (Fig.~\ref{fig:1}a; see also Supplementary Section~\ref{sec:fabrication}). The optical and mechanical degrees of freedom are coupled by radiation pressure, whereby intracavity photons exert a force on the cavity boundaries and the resulting mechanical displacement shifts the optical resonance frequency~\cite{Aspelmeyer2014}. In silicon, however, radiation pressure is not the only nonlinear mechanism at play: at the intracavity photon numbers required to reach self-sustained oscillation, two-photon absorption generates free carriers that induce both free-carrier dispersion and absorption, while carrier relaxation produces heating that activates a thermo-optic shift of the cavity resonance~\cite{Barclay2005}. These processes are dynamically coupled through the intracavity field and unfold on distinct timescales---from the sub-nanosecond optical response to nanosecond-scale carrier and thermal relaxation---forming a multi-timescale nonlinear oscillator. Such interplay between optomechanical motion, two-photon absorption, free-carrier and thermo-optic dynamics has been shown to produce self-pulsing, bistability and chaos in silicon optomechanical cavities~\cite{NavarroUrrios2017}. All these dynamical regimes are accessible through the effective detuning between the laser and the optical cavity resonance, $\Delta$, or by the input laser power, $P_\mathrm{in}$.

In the self-pulsing regime, the intracavity field undergoes autonomous nonlinear oscillations that generate a strongly anharmonic modulation of the photon number; through radiation pressure, this modulation acts as an internally generated drive with spectral components at the self-pulsing frequency and its harmonics~\cite{NavarroUrrios2016}. When one of these components approaches the mechanical resonance frequency, it provides an effective periodic forcing that entrains the mechanical motion and leads to coherent self-sustained oscillation. This mechanism differs from conventional radiation-pressure-driven dynamical backaction, where mechanical amplification originates from a feedback process in which mechanical motion modulates the intracavity field, and the resulting radiation-pressure force---delayed by the cavity lifetime---modifies the mechanical dynamics~\cite{Aspelmeyer2014}; our system reaches a hybrid optomechanical limit cycle in which optical nonlinearities associated with self-pulsing provide the effective drive, while optomechanical feedback selects the oscillation frequency and phase and stabilizes the coherent motion. This resulting phonon-lasing regime can be accessed even in the unresolved-sideband limit characteristic of low-optical-quality($Q$)-factor cavities which offers two key advantages. First, the broader optical linewidth exceeds the device-to-device cavity-frequency spread imposed by fabrication disorder~\cite{Minkov2013}. This allows multiple nominally identical (by design) cavities to be addressed simultaneously with a shared optical drive---whereas resolved-sideband platforms require unrealistic ultra-low-disorder fabrication to secure spectral overlap for simultaneous driving. Second, high optical $Q$ restricts the spectral bandwidth of the optical signal, whereas the low-$Q$ regime supports a broader spectral range for information encoding.

As the detuning brings the laser onto the cavity resonance, the broad thermal noise peak sharpens into a coherent mechanical tone, marking the thermally broadened onset of the stable limit cycle---phonon lasing (Figs.~\ref{fig:1}b--d). Above threshold, the cavity acts as a single nonlinear node that implements the reservoir: the input laser plays two distinct roles simultaneously, with its constant component $\bar{P}_\mathrm{in}$ driving the limit cycle while a weak modulation $\delta P_\mathrm{in}$ around this operating point encodes the computational input $s(t)$, i.e., the information we need to solve a given task. This modulation deforms the nonlinear attractor while preserving the limit cycle, and the evolving cavity state maps the input onto a nonlinear output $y(t)=f_\mathrm{NL}[s(t)]$---the transmitted optical power (Fig.~\ref{fig:1}e)---whose amplitude and phase are set by the coupled optomechanical dynamics. This output is recorded and used as the reservoir state from which a linear readout is trained: for a chosen task, we provide a set of inputs $s(t)$ together with the known desired outputs $o(t)$ — the targets — and the readout weights are fit so that the readout reproduces $o(t)$ from the reservoir state. Depending on the task, the target can be a nonlinear function of the input, a future value of a time series, or a class label. Once trained, the readout can be applied to new inputs to produce the corresponding outputs. The nonlinearity of the cavity response is essential: a linear node would map the modulation onto a proportional output, and the linear readout could not perform any computation beyond a linear transformation of the input. This encoding requires a perturbative regime that preserves the stability of the limit cycle, in which the computation arises from controlled deformations of the attractor; only here does the cavity retain the nonlinearity, dynamical memory, and reproducibility required for a stable input--output mapping. Larger modulation amplitudes drive the system away from the limit cycle, disrupting the dynamics and suppressing the computational capability. Conversely, below threshold there is no limit cycle to perturb. Experimentally, we find that the Brownian regime cannot serve as a reservoir: modulating a stochastic signal produces a noise-dominated output that lacks the reproducible correlations required for information processing~\cite{Wiesner2021}, and no reliable input--output mapping can be extracted (Supplementary Fig.~S5a).

A natural approach for reservoir computing in this system is time multiplexing~\cite{Appeltant2011}, which involves two conceptual steps. First, the computational task is encoded by mapping a discrete input sequence $\{s(k)\}$, with corresponding targets $\{o(k)\}$, onto the continuous-time modulation of the laser intensity $\delta P_\mathrm{in}(t)$ that drives the cavity. Second, the input data is time-multiplexed: each input value $s(k)$ is multiplied by a random mask with $N$ discrete levels, and each masked value is held for a duration $\tau$ (the node duration), so that the total time allocated to one input step is $T_\mathrm{in} = N\tau$. The $N$ samples of the cavity response within each input step then play the role of $N$ virtual nodes (Fig.~\ref{fig:1}e; see also Supplementary Section~\ref{sec:timemux}). This dimensional expansion is computationally meaningful only because the cavity is nonlinear: each masked perturbation drives the attractor along a different direction, so that the $N$ transient responses within one input step span a high-dimensional state space. A linear node would simply superpose the masked inputs, and all $N$ responses would collapse onto a single linear combination of $s(k)$, providing no dimensional gain. The linear readout described above is then trained on this high-dimensional reservoir state. The richness of this representation can be directly assessed by testing the ability of the system to reconstruct nonlinear functions. Using input data drawn from a uniform distribution and defining the target as $o(k)=f(s(k))$, we show in Fig.~\ref{fig:2} that, with only 20 virtual nodes, the reservoir accurately reconstructs $\sin(s(k))$, $\cos(s(k))$, $s(k)^2$, and $s(k)^3$. Performance is quantified using the normalized mean square error (NMSE), defined as the mean squared deviation between prediction and target normalized by the variance of the target signal. At $N=20$ virtual nodes and $T_\mathrm{in}=72$~ns ($\tau=3.6$~ns per node), we obtain NMSE~$\approx 0.1$ for $\sin(s)$ and NMSE~$\approx 0.08$ for $s^3$, compared with NMSE~$\approx 0.13$ for $\cos(s)$ and for $s^2$. While these NMSE values reflect both the intrinsic cavity response and the statistical structure of the targets, odd functions are systematically better reconstructed across all operating conditions. The even/odd gap is most pronounced at small node numbers: at $N=4$, odd functions already reach near-optimal performance while even functions remain above NMSE~$\approx 0.8$, indicating that odd targets are well captured even by a low-dimensional embedding. Both function classes improve with increasing $\tau$, consistent with the node duration approaching the mechanical period and allowing each virtual node to resolve a full nonlinear oscillation cycle (Supplementary Fig.~S10). Function reconstruction is a memoryless task --- the target $o(k) = f(s(k))$ depends only on the current input --- so the performance improvement with $\tau$ reflects richer nonlinear expansion of the attractor rather than any memory effect. The reconstruction accuracy and the odd/even asymmetry together characterize the nonlinear response of the cavity around the phonon-lasing operating point as predominantly low-order, consistent with weak perturbations of the coherent attractor.

Temporal processing requires that the reservoir state depend on recent inputs while progressively losing dependence on distant ones. In the coherent regime above the Hopf bifurcation, this fading memory arises intrinsically from the coexistence of multiple dynamical timescales: self-sustained oscillations persist over many mechanical periods ($T_m = 2.39$~ns), while free-carrier and thermo-optic relaxation introduce slower dynamics with a characteristic lifetime of $\sim$10~ns. The instantaneous cavity state therefore encodes a temporally weighted history of recent masked inputs. The linear memory capacity quantifies how well this physical memory can be accessed by a linear readout to reconstruct past input values,
\begin{equation}
\mathrm{MC} = \sum_{p=1}^{P} \mathrm{corr}(s(k-p), \hat{o}_p(k)),
\end{equation}
where $\hat{o}_p(k)$ is the reservoir's reconstruction of the input $p$ steps in the past and $\mathrm{corr}(\cdot,\cdot)$ denotes the Pearson correlation (see Supplementary Information). Figure~\ref{fig:3}a shows measured fading-memory diagrams for representative values of the memory capacity: in all cases the correlation is strongest at $|p|=1$ and decays for increasing delay, confirming the fading-memory property, with the decay rate determining the accessible memory depth. For $\mathrm{MC} \approx 0.98$ correlations vanish at $|p|\geq2$, whereas for $\mathrm{MC} \approx 3$ significant correlations persist up to $|p|\sim3$--4, enabling processing of temporal sequences requiring multiple past inputs. The memory capacity increases with node duration $\tau$ (Fig.~\ref{fig:3}b), as longer nodes allow integration over multiple mechanical cycles, while performance converges with modest node number (Fig.~\ref{fig:3}c), indicating that memory rather than dimensionality limits performance in this regime. Numerical simulations of the validated coupled-mode model (Supplementary Fig.~S11) show that the attainable memory capacity is governed by the separation between fast and slow timescales as well as by the dynamical regime of the oscillator.
The coexistence of two dynamical timescales and optimizing the relation between them (by changing the value of $\Omega_m$ and by selecting a specific dynamical regime), enables memory capacity of three past steps.
Because the thermal relaxation time is set by material properties and largely independent of geometry, while $\Omega_m$ is defined by the device dimensions at fabrication, the maximum memory depth becomes a design parameter set by the mechanical frequency rather than tuned at runtime, in contrast to delay-line reservoirs where memory is implemented through an external feedback path with an independent timescale. In this system, dissipation plays a dual role: the finite relaxation times associated with carrier and thermal dynamics enable temporal integration and memory formation, while the same dissipative processes limit memory by driving the decay of correlations, thereby setting the intrinsic memory horizon.

Nonlinearity and memory are the essential ingredients for temporal information processing. We test both properties simultaneously on the prediction of the Mackey–Glass chaotic time series~\cite{Mackey1977}, whose weakly chaotic dynamics require nonlinear transformation together with short-term memory for $p$-step-ahead prediction. Figure~\ref{fig:4}a shows the normalized mean square error (NMSE) as a function of the node duration $\tau$ for one-step ($p=1$) and two-step ($p=2$) prediction using 20 virtual nodes. The error decreases with increasing $\tau$ and reaches a minimum within the explored range, indicating that longer node durations --- spanning multiple mechanical periods and approaching the thermal relaxation time --- enable the cavity state to integrate information over more oscillation cycles and thereby access deeper memory. The clear separation between $p=1$ and $p=2$ performance reflects the greater difficulty of the $p=2$ task and the finite predictability horizon set by dissipation, as mechanical damping and thermal relaxation progressively reduce correlations with earlier states. Figure~\ref{fig:4}b further shows that increasing the number of virtual nodes $N$ improves performance up to a modest reservoir size, beyond which the gain saturates, indicating that the limiting factor is not dimensionality but the effective memory capacity of the system. Representative time traces in Figs.~\ref{fig:4}c and~\ref{fig:4}d show that for $p=1$ the reservoir reproduces both amplitude and phase of the chaotic oscillation with high fidelity, whereas for $p=2$ deviations gradually become more pronounced as the prediction approaches the system’s predictability horizon. The degradation in performance is therefore not algorithmic but dynamical, directly governed by the interplay between mechanical oscillation, thermal response, and the node duration that controls how strongly past states influence the evolving cavity dynamics.

\begin{figure}[t]
\centering
\includegraphics[width=0.4\textwidth]{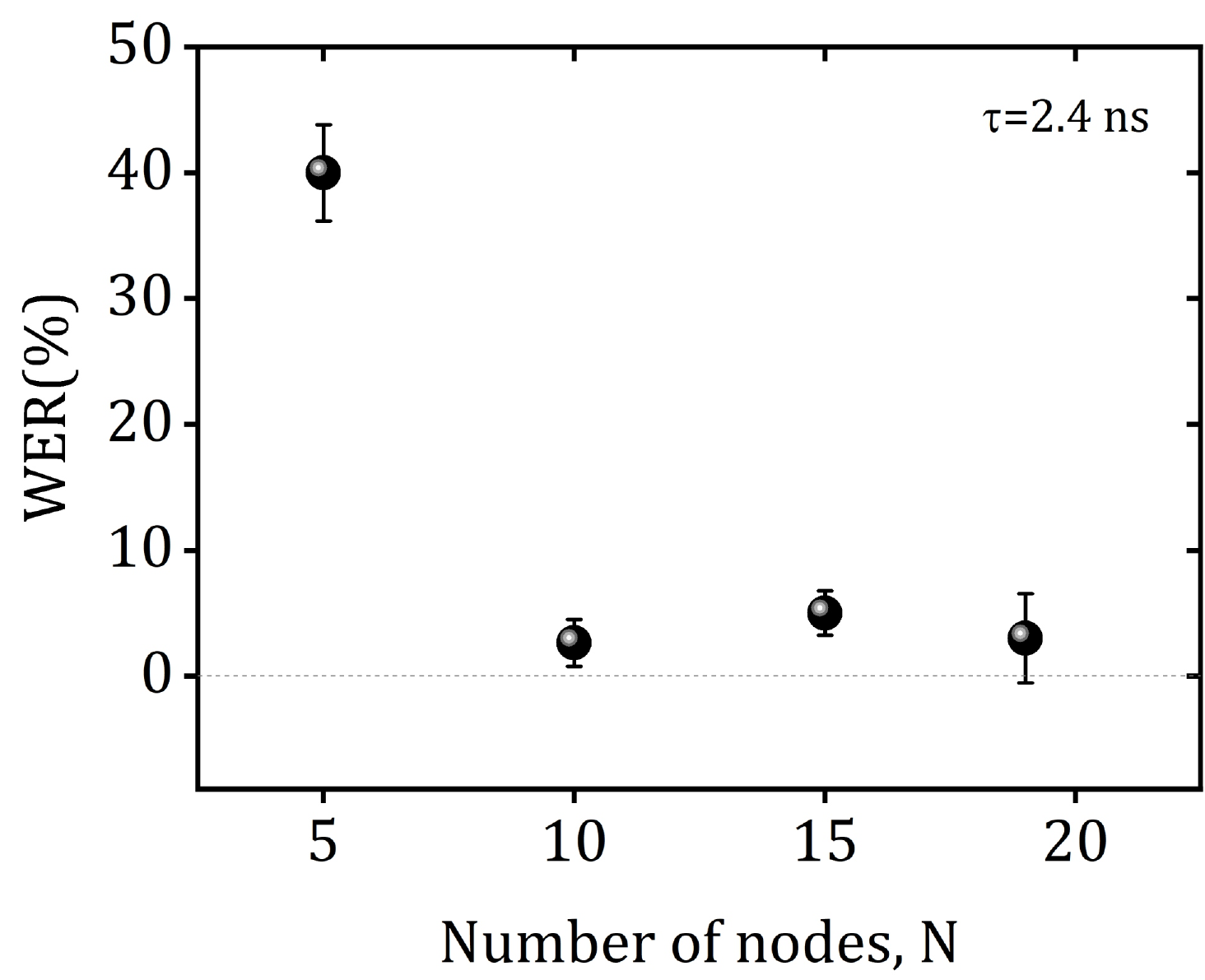}
\caption{%
\textbf{Spoken digit recognition.}
Word error rate (WER) as a function of the number of virtual nodes $N$ for a two-digit classification task using the TI-46 speech corpus. Performance converges to near-zero WER at modest node numbers ($N \geq 10$), consistent with the memory-limited rather than dimensionality-limited operation of the reservoir. Error bars indicate the standard deviation over repeated measurements.}
\label{fig:5}
\end{figure}

To further benchmark the computational capability of the reservoir on a real-world classification task, we tested spoken digit recognition using the TI-46 speech corpus~\cite{TI46}, a standard evaluation benchmark for reservoir computing~\cite{Verstraeten2005}. Each spoken digit is preprocessed using the Lyon cochlear model, which converts the raw acoustic waveform into an 86-channel frequency--time cochleagram representing the spectral power distribution over the duration of the utterance. The cochleagram is projected onto the reservoir using a two-dimensional random mask of dimensions $N \times 86$, producing a one-dimensional time-multiplexed input that is injected into the cavity with the same hold-time protocol used for the Mackey--Glass task. The linear readout is trained against a target matrix encoding which digit was spoken. The 4~$\mu$s coherent oscillation window of the present device limits the number of utterances that can be processed within a single coherent run (see Supplementary Information), so we evaluate performance on a two-digit classification sub-task; extending this window through higher-$Q$ cavities or active phase stabilization would enable the full ten-class task. Figure~\ref{fig:5} shows the word error rate (WER) as a function of the number of virtual nodes $N$. Performance improves sharply between $N=5$ and $N=10$, where the WER drops from $\sim$40\% to near zero, and remains low for larger $N$. This rapid convergence indicates that the reservoir provides sufficient nonlinear expansion to linearly separate the two classes with a relatively small number of virtual nodes, while further increases in dimensionality yield diminishing returns. These results show the capability of the system to process structured temporal inputs.

The tasks reported here are ultimately limited by the dynamical timescale of the reservoir, which in optomechanical systems is set directly by the mechanical resonance frequency, which governs how rapidly the system state can evolve and therefore the processing rate of the reservoir. In the present device, operating near 0.4~GHz, this corresponds to nanosecond-scale node durations --- a proof-of-concept that establishes the underlying physics while operating well below the highest frequencies accessible in related nanomechanical platforms. That range is not fixed: advances in nanostructured phononic systems have confined mechanical resonances from the MHz regime to multi-GHz frequencies on the same silicon-on-insulator platform~\cite{Florez2022}. Because the reservoir dynamics follow the mechanical motion directly, increasing the mechanical frequency leads to a proportional increase in computational throughput, identifying optomechanical reservoir computing as a platform with a clear scaling path toward high-speed physical information processing.

We have experimentally accessed one such dynamical boundary --- the thermally-broadened Hopf bifurcation from stochastic Brownian motion to coherent phonon lasing --- and shown that the limit cycle above threshold simultaneously provides nonlinearity, memory, and reproducibility to solve different reservoir-computing tasks. More broadly, this optomechanical oscillator realises a physical instance of a paradigm proposed independently in computational neuroscience and machine learning: that of a damped harmonic oscillator coupled to an information-carrying field~\cite{effenberger2025functional, ichikawa2021short}. This same physical system reaches further dynamical regimes: chaotic dynamics have been observed in single nanocavities~\cite{NavarroUrrios2017}, and synchronization between mechanically coupled cavities has been also shown in optomechanical systems~\cite{Colombano2019}. Both regimes are computationally advantageous: information-processing capacity is often maximized near critical boundaries, either locally at the node level, such as bifurcations and the edge of chaos~\cite{Akashi2020,Bertschinger2004}, or collectively through synchronization phase transitions~\cite{diSanto2018}. Coupled optomechanical oscillator networks, with individually tunable bifurcation parameters and engineered inter-cavity couplings, offer a physical platform to study how local and collective criticality together govern information processing.

\vspace{8pt}
{\small\bfseries Acknowledgments.}
{\small This work was supported by the HORIZON-EIC-2022
Pathfinder project NEUROPIC (Grant No.\ 101098961),
the Spanish Ministry of Science, Innovation and Universities
via the national project PID2024-158832NB-C21 (PSYNC),
and the Spanish MICIU Severo Ochoa program for Centers
of Excellence (Grant CEX2024-001445-S). We acknowledge
Dr.~G.~Arregui for valuable discussions.}



\clearpage
\onecolumn
\fontsize{12pt}{13.5pt}\selectfont

\begin{center}
{\large\bfseries SUPPLEMENTARY INFORMATION}
\end{center}
\vspace{1em}

\setcounter{figure}{0}
\setcounter{table}{0}
\setcounter{equation}{0}
\renewcommand{\thefigure}{S\arabic{figure}}
\renewcommand{\thetable}{S\Roman{table}}
\renewcommand{\theequation}{S\arabic{equation}}

\titleformat{\section}{\normalsize\bfseries\centering\MakeUppercase}{}{0em}{}
\titleformat{\subsection}{\normalsize\bfseries\centering}{}{0em}{}
\titlespacing{\section}{0pt}{18pt}{10pt}
\titlespacing{\subsection}{0pt}{12pt}{6pt}

\setcounter{topnumber}{1}
\setcounter{bottomnumber}{1}
\setcounter{totalnumber}{2}
\renewcommand{\topfraction}{0.85}
\renewcommand{\bottomfraction}{0.85}
\renewcommand{\textfraction}{0.15}
\renewcommand{\floatpagefraction}{0.85}

\section{Sample design and fabrication}
\label{sec:fabrication}

The optomechanical cavity consists of two photonic crystal slabs ($30~\mu\mathrm{m} \times 4.5~\mu\mathrm{m} \times 220~\mathrm{nm}$) separated by a 55~nm wide slot across the photonic crystal line defect. The hole radius is $r = 0.32a$ with a lattice constant $a = 445$~nm. To form a localized cavity resonance, the local lattice constant is smoothly reduced from the nominal value of 445~nm to 435~nm in the center of the cavity. This design produces a confined optical mode with a resonance wavelength of 1558.6~nm and a measured quality factor $Q \approx 60{,}000$. The slab dimensions determine the in-plane mechanical mode frequency, which is $\Omega_m / 2\pi = 418$~MHz (period $T_m = 2.39$~ns). The vacuum optomechanical coupling rate between the confined optical mode and the extended mechanical mode is measured as $g_0/2\pi = 551.9 \pm 19.1$~kHz via the optomechanical backaction method and independently confirmed as $g_0/2\pi = 709.2 \pm 54.7$~kHz via phase-modulation calibration (see Section~\ref{sec:characterization}).

The cavities are fabricated on a silicon-on-insulator (SOI) substrate consisting of a 220-nm silicon device layer separated from a 775-$\mu$m silicon handle by a 2-$\mu$m buried oxide (BOX) layer. The cavity pattern is defined in a 180-nm-thick ZEP 530-A resist by high-resolution electron-beam lithography with a 1-nm shot pitch, which minimises line-edge roughness. The patterned resist serves as a mask for anisotropic plasma dry etching, transferring the design into the silicon device layer. The slabs are then mechanically released by selectively removing the BOX beneath the cavity using vapor-phase hydrofluoric acid (HF) etching, which avoids the stiction associated with surface tension in a liquid-phase release. The isotropic chemical etch forms a $\sim$2-$\mu$m lateral undercut in the BOX measured from the edge of the silicon device. Figure~\ref{fig:S_fab} shows top-view scanning electron microscope (SEM) images of the fabricated cavity at two magnifications, resolving the periodic hole array, the central slot that defines the line defect, and the smooth sidewall profiles of the etched features.

\setcounter{figure}{-1} 
\begin{figure}[htbp]
    \centering
    \includegraphics[width=\textwidth]{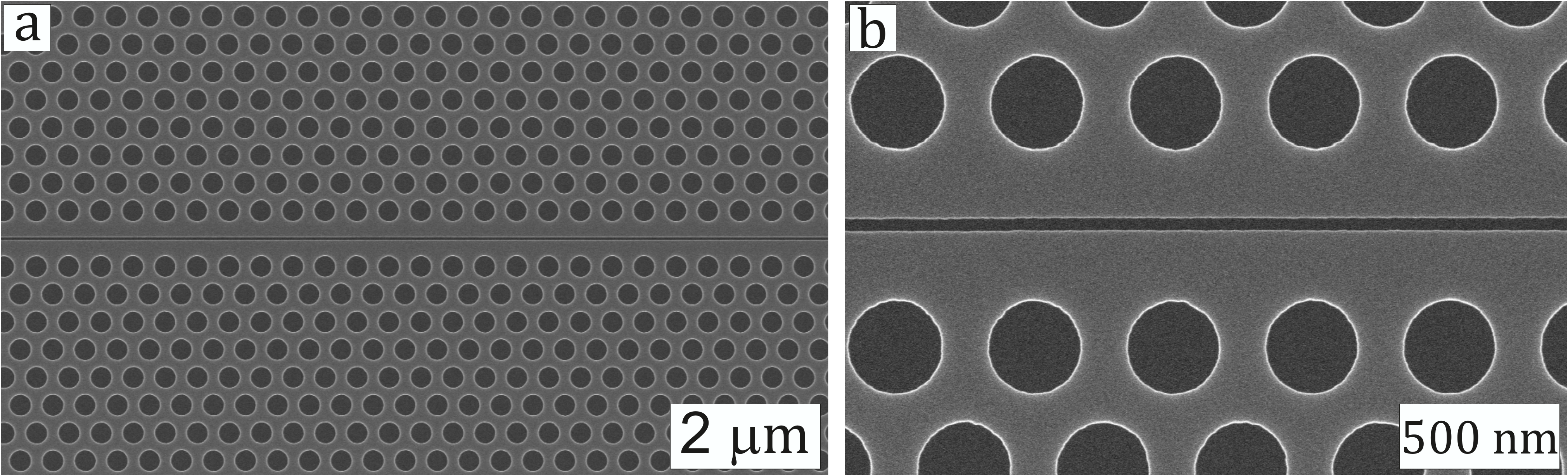}
    \caption{Scanning electron microscope (SEM) images of the fabricated photonic crystal cavity. (\textbf{a})~Top view of the device showing the periodic hole array on both sides of the central slot that defines the line defect. (\textbf{b})~Close-up of the slot region, resolving the $\sim$55-nm slot separating the two photonic crystal slabs and the smooth, anisotropic profiles of the etched holes. Scale bars: 2~$\mu$m in (a), 500~nm in (b).}
    \label{fig:S_fab}
\end{figure}

\section{Model of the silicon optomechanical cavity dynamics}
\label{sec:model}

The dynamics of the optomechanical photonic crystal cavity are modeled within the time-domain nonlinear coupled-mode formalism, incorporating two primary physical mechanisms: optomechanical coupling~\cite{Aspelmeyer2014_SI} and two-photon absorption (TPA)~\cite{Johnson:06,mesochaos}, from which two further effects derive --- free-carrier dispersion and thermo-optic frequency shifts. The system is described by the following set of four coupled differential equations.

The optomechanical equation of motion describes the optically driven damped mechanical harmonic oscillation:
\begin{equation}\label{eq:S_mech}
    \frac{d^2x}{dt^2} + \Gamma_m \frac{dx(t)}{dt} + \Omega_m^2 x(t) = \frac{g_0}{\omega_0} \sqrt{\frac{2\Omega_m}{\hbar\, m_\mathrm{eff}}} |U(t)|^2.
\end{equation}
Here $x(t)$ is the mechanical displacement due to radiation pressure, $\Gamma_m$ is the mechanical damping rate, $\Omega_m$ is the mechanical resonance frequency, $g_0$ is the vacuum optomechanical coupling rate, $\omega_0$ is the optical cavity resonance frequency, and $m_\mathrm{eff}$ is the effective motional mass. The term on the right-hand side represents the radiation-pressure force proportional to the intracavity photon energy $|U(t)|^2$.

The intracavity optical field evolves according to:
\begin{align}\label{eq:S_optical}
    \frac{dU}{dt} &= i\biggl(-g_0 \sqrt{\frac{2 m_\mathrm{eff} \Omega_m}{\hbar}} x(t) + \frac{\omega_0}{n_\mathrm{si}} \left(\frac{dn_\mathrm{si}}{dT}\Delta T(t) + \frac{dn_\mathrm{si}}{dN} N(t)\right) + \delta\omega \biggr) U(t) \nonumber \\
    &- \frac{1}{2}\left(\gamma_i + \gamma_e + \frac{\Gamma_\mathrm{TPA}\,\beta_\mathrm{si}\, c^2}{V_\mathrm{TPA}\, n_g^2} |U(t)|^2 + \frac{\sigma_\mathrm{si}\, c\, N(t)}{n_g}\right) U(t) \nonumber \\
    &+ \sqrt{\gamma_e \left(\bar{P}_\mathrm{in} + \delta P_\mathrm{in}\, U_\mathrm{inj}(t)\right)},
\end{align}
where $\delta\omega = \omega_L - \omega_0$ is the detuning between the input laser frequency and the cavity resonance, $\gamma_i$ and $\gamma_e$ are the intrinsic and extrinsic optical decay rates, $n_g$ is the group index, $c$ is the speed of light, $\beta_\mathrm{si}$ is the TPA coefficient of silicon, and $\sigma_\mathrm{si}$ is the free-carrier absorption cross section. The input power $\bar{P}_\mathrm{in} + \delta P_\mathrm{in}\, U_\mathrm{inj}(t)$ includes a constant DC component $\bar{P}_\mathrm{in}$ and an optional modulation component $\delta P_\mathrm{in}$. The terms $dn_{si}/dT$ and $dn_{si}/dN$ account for thermo-optic and free-carrier dispersion shifts of the cavity resonance, proportional to changes in temperature and number of free carriers respectively.

The free-carrier density is governed by TPA generation and recombination:
\begin{equation}\label{eq:S_fc}
    \frac{dN}{dt} = -\gamma'_\mathrm{fc}\, N(t) + \frac{\Gamma_\mathrm{fc}\,\beta_\mathrm{si}\, c^2}{2\hbar\omega_p\, n_g^2\, V_\mathrm{FCA}^2} |U(t)|^4,
\end{equation}
where $\gamma'_\mathrm{fc}$ is the free-carrier recombination rate and the second term describes the TPA-generated carrier density, with $\Gamma_\mathrm{fc}$ and $V_\mathrm{FCA}$ the confinement factor and effective mode volume for free-carrier absorption, respectively.

The cavity temperature variation follows:
\begin{align}\label{eq:S_thermal}
    \frac{d\Delta T}{dt} &= -\gamma_\mathrm{th}\,\Delta T(t) + \frac{\Gamma_\mathrm{phc}}{\rho_\mathrm{si}\, C_p\, V_\mathrm{phc}} \left(\gamma_\mathrm{lin} + \frac{\Gamma_\mathrm{TPA}\,\beta_\mathrm{si}\, c^2}{V_\mathrm{TPA}\, n_g^2} |U(t)|^2 + \frac{\sigma_\mathrm{si}\, c}{n_g} N(t)\right) |U(t)|^2,
\end{align}
where $\gamma_\mathrm{th}$ is the thermal dissipation rate (corresponding to a thermal lifetime of $\sim 10$~ns), $\rho_\mathrm{si}$ is the silicon density, $C_p$ is the heat capacity, and $V_\mathrm{phc}$ is the photonic crystal mode volume. The three terms inside the parentheses represent heating from linear absorption, TPA, and free-carrier absorption, respectively.

The nonlinear response of the system arises primarily from radiation-pressure coupling and intensity-dependent refractive index changes induced by free carriers and thermo-optic effects. Memory emerges from the coexistence of multiple dynamical timescales: fast mechanical oscillations at hundreds of MHz, intermediate free-carrier dynamics, and slower thermal relaxation. The interplay between the mechanical oscillation period ($T_m = 2.39$~ns at $\Omega_m / 2\pi = 418$~MHz) and the thermal dissipation lifetime ($\sim 10$~ns) is particularly central to the reservoir's memory properties, as discussed in detail in Section~\ref{sec:timemux}. In the time-multiplexed implementation, the mechanical oscillation defines the natural clock of the reservoir, while the slower carrier and thermal dynamics introduce temporal correlations between successive virtual nodes, providing the fading memory required for reservoir computing. These virtual nodes correspond to discrete time samples of the system's response within each input period, effectively expanding a single nonlinear node into a high-dimensional state space~\cite{Appeltant2011_SI}. A complete list of the simulation parameters is provided in Ref.~\cite{mesochaos}.

\section{Time-multiplexed reservoir computing}
\label{sec:timemux}

Reservoir computing is a framework for computation with dynamical systems in which a fixed, nonlinear system --- the reservoir --- maps input signals into a high-dimensional state space, and only a simple linear readout layer is trained~\cite{Appeltant2011_SI}. The key requirements are nonlinearity and fading memory: the reservoir must transform its inputs in a nonlinear way and retain information about past inputs over a finite time horizon. In the time-multiplexed approach, a single physical node is used to emulate a large network of virtual nodes by injecting successive masked versions of the input and sampling the node response at different times~\cite{Appeltant2011_SI}. This strategy is only useful if the physical node is nonlinear: a linear node would simply superpose the masked inputs without performing any nontrivial transformation, and the virtual nodes would carry no independent computational information. The optomechanical cavity operated in the phonon-lasing regime provides precisely the required nonlinearity through radiation-pressure coupling, free-carrier dispersion, and thermo-optic effects, as described in Section~\ref{sec:model}.

\subsection{Input data preparation}

\begin{figure}[htbp]
    \centering
    \includegraphics[width=0.8\textwidth]{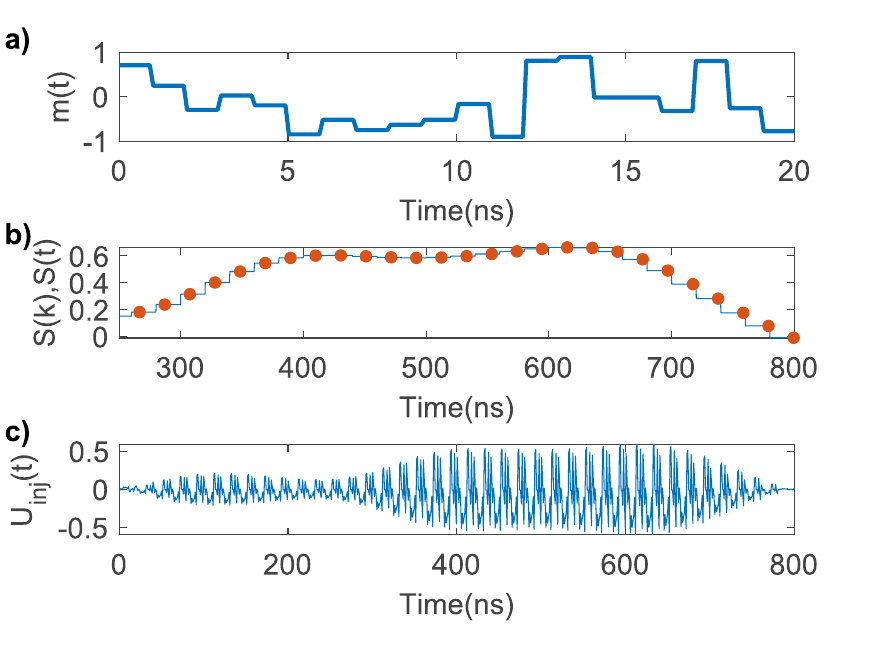}
    \caption{Time-multiplexed input signal generation. (a)~Randomly drawn mask $m(t)$ generating $N = 20$ virtual nodes, where each step has a duration $\tau$ (the hold time). (b)~Discrete input data $s(k)$ (red circles) converted to a continuous signal $s(t)$ by holding each data point constant for a total period $N\tau$. (c)~Time-multiplexed input sequence $U_\mathrm{inj}(t) = s(t)\, m(t)$.}
    \label{fig:S_timemux}
\end{figure}

To generate a large number of virtual nodes using a single optomechanical cavity, the input data is time-multiplexed~\cite{Appeltant2011_SI}. The procedure is illustrated in Fig.~\ref{fig:S_timemux}. A mask consisting of $N$ random values drawn from a uniform distribution is generated. Each input data point $s(k)$, $k \in \{1, \ldots, K\}$, is multiplied by the $N$ mask values, producing $N \times K$ distinct input values. Since the reservoir operates continuously in time, each discrete masked input value is held for a duration $\tau$, generating a continuous time-multiplexed input signal:
\begin{equation}\label{eq:S_input}
    U_\mathrm{inj}(t) = s(t)\, m(t),
\end{equation}
where $\tau$ is the hold time (the duration assigned to each virtual node) and $N$ is the number of virtual nodes, so that the total time allocated to one input data point is $N\tau$. The time-multiplexed signal is used to modulate the amplitude of the cavity input signal. For all tasks reported in this work, $K = 3{,}000$ data points were used.

\subsection{Readout and training}

The signal transmitted from the cavity is recorded by the oscilloscope as a sequential time trace. For each discrete masked input value, a corresponding discrete output value is extracted by averaging the output signal over the equivalent input period, yielding $K \times N$ discrete output values. These are divided into a training set (2{,}000 points) and a test set (1{,}000 points).

The training output values are reshaped into a states matrix:
\begin{equation}\label{eq:S_states}
X_\mathrm{train} = \begin{pmatrix}
x_1(1) & \cdots & x_1(K_\mathrm{train}) \\
\vdots & \ddots & \vdots \\
x_N(1) & \cdots & x_N(K_\mathrm{train})
\end{pmatrix},
\end{equation}
where $x_n(k)$ denotes the output of virtual node $n$ at time step $k$. The predicted output is obtained from the states matrix as $\hat{o}(k) = W\, X_\mathrm{train}$, where $W$ is a weight vector of dimension $1 \times N$ determined by linear regression (ridge regression) that minimizes the error between the predicted output $\hat{o}(k)$ and the target $o(k)$. Once $W$ is obtained from the training data, the test set is used to evaluate the normalized mean square error:
\begin{equation}\label{eq:S_NMSE}
    \mathrm{NMSE} = \frac{1}{K_\mathrm{test}} \sum_{k=1}^{K_\mathrm{test}} \frac{\left(o(k) - \hat{o}(k)\right)^2}{\sigma^2\!\left(o(k)\right)},
\end{equation}
where $\sigma^2(o(k))$ is the variance of the target data over the test set.

\begin{figure}[htbp]
    \centering
    \includegraphics[width=0.8\textwidth]{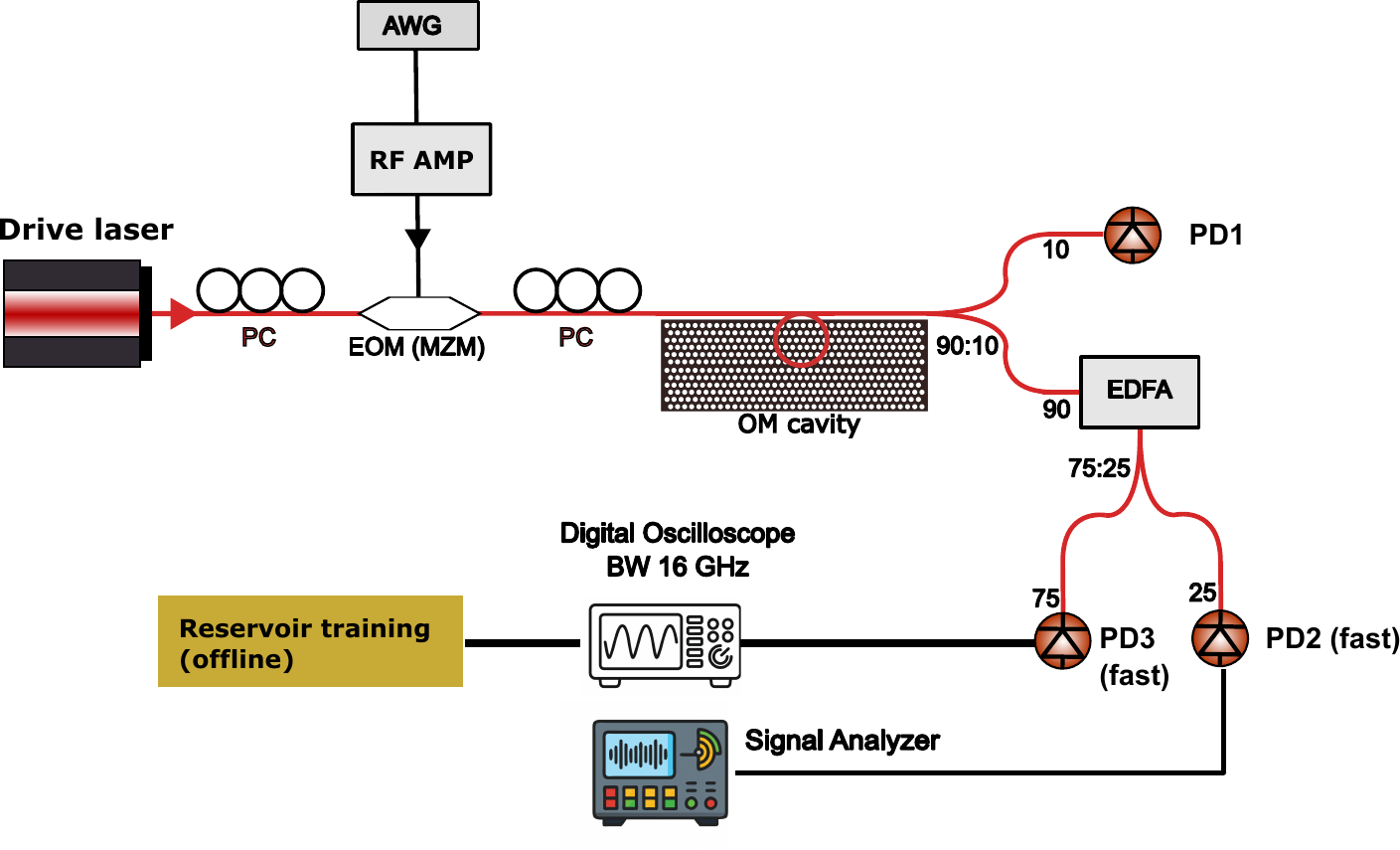}
    \caption{Schematic of the reservoir computing experimental setup. The input laser signal is passed through an intensity Mach-Zehnder modulator (MZM) to encode the reservoir input generated by the arbitrary waveform generator (AWG) into the cavity. The signal is coupled into the optomechanical cavity via the evanescent field of a tapered fiber loop. The transmitted signal is amplified by an erbium-doped fiber amplifier (EDFA) and split between photodetectors: 25\% is sent to a signal analyzer to monitor the self-sustained oscillation, and 75\% is recorded by an oscilloscope for offline reservoir training. AWG, arbitrary waveform generator; PC, polarization controller; PD, photodiode.}
    \label{fig:S_setup}
\end{figure}

\section{Experimental setup}

A schematic of the experimental setup is shown in Fig.~\ref{fig:S_setup}. A tunable continuous-wave laser (Santec TSL-570) is connected to a lithium-niobate Mach-Zehnder intensity modulator that encodes the reservoir input data onto the optical carrier. The modulated light is coupled into the optomechanical cavity via the evanescent field of a tapered fiber loop positioned in contact with the nanophotonic structure.

The input data signals are generated by an arbitrary waveform generator (Rigol DG70000) connected to the Mach-Zehnder modulator. The time-multiplexed signal $U_\mathrm{inj}(t)$ described in Section~\ref{sec:timemux} is loaded into the AWG, which drives the modulator to encode the masked input sequence onto the optical carrier. The transmitted cavity signal is amplified by an erbium-doped fiber amplifier (EDFA) and split between two high-speed photodetectors. A fraction of the signal (25\%) is sent to a 20~GHz signal analyzer (Signal Hound SM200B), which serves as a real-time probe to verify that the cavity maintains self-sustained optomechanical oscillations throughout the measurement. The main portion of the signal (75\%) is sent to a 16~GHz bandwidth oscilloscope (Rohde \& Schwarz RTP164B) from which the data is downloaded for offline processing.

The wavelength of the input optical signal is detuned from the cavity resonance. The detuning and input power are chosen such that the optomechanical cavity operates in a steady-state regime of self-sustained oscillations at a fixed amplitude and at the frequency of the fundamental mechanical mode. We measured a maximum coherent oscillation time of $4~\mu$s during which the self-sustained oscillations are maintained without significant phase noise, allowing multiple measurements to be averaged. This averaging is necessary to reduce noise and is crucial for good reservoir performance. The input signal is divided into $4~\mu$s segments, each measured multiple times and averaged by the oscilloscope. The physical origin of this coherence limitation is not addressed here and will be the subject of future work.

\section{Optomechanical characterization}
\label{sec:characterization}

\begin{figure}[htbp]
    \centering
    \includegraphics[width=0.95\textwidth]{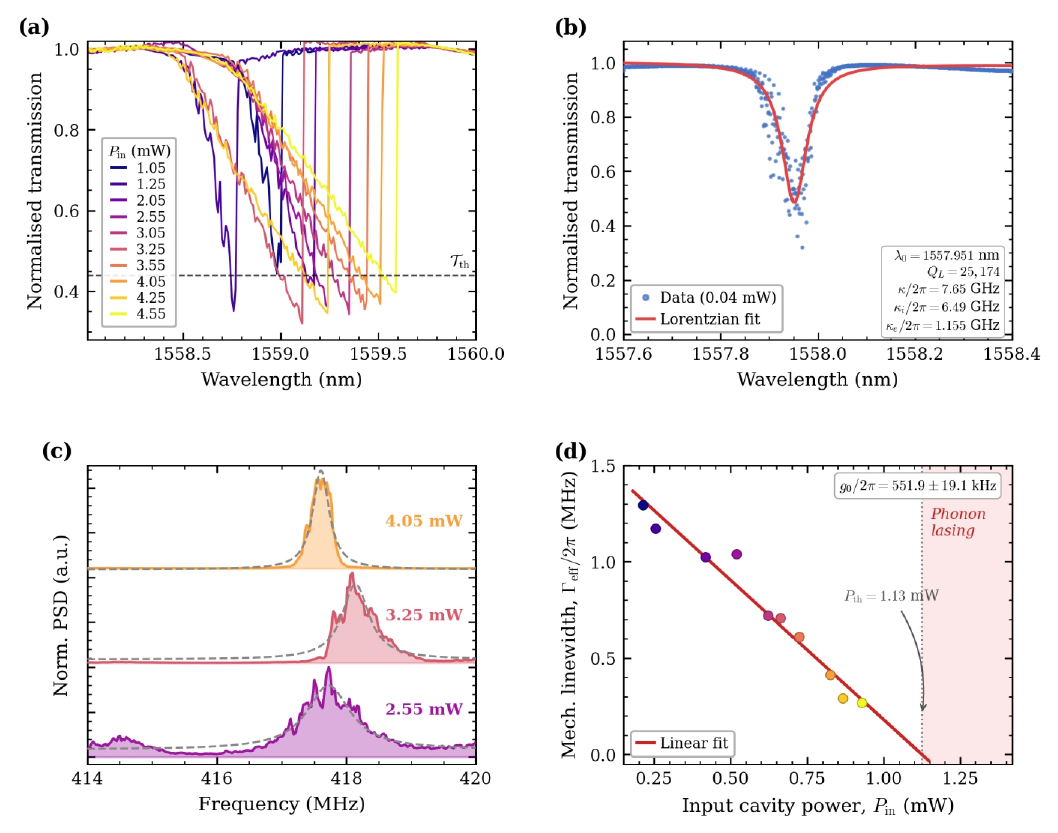}
    \caption{Optomechanical characterization of the photonic crystal cavity.
    (\textbf{a})~Normalised optical transmission spectra recorded as a function of
    input laser wavelength for ten input powers $P_\mathrm{in}$ between 1.05 and
    4.55~mW.  The progressive red-shift of the resonance dip with increasing power
    reflects thermo-optic heating of the silicon slab.  The dashed horizontal line
    marks the threshold transmission $\mathcal{T}_\mathrm{th} \approx 0.44$ used to
    select the operating detuning $\Delta \approx -\kappa/2$ at each power.
    (\textbf{b})~Normalised optical transmission at low power (0.04~mW, blue dots)
    and Lorentzian fit (red curve), yielding $\lambda_0 = 1557.951$~nm,
    $\kappa/2\pi = 7.65$~GHz, $\kappa_i/2\pi = 6.49$~GHz,
    $\kappa_e/2\pi = 1.155$~GHz, and $Q_L = 25{,}174$.
    (\textbf{c})~Normalised mechanical power spectral density (PSD) spectra at
    three input powers (2.55, 3.25, and 4.05~mW), measured at the blue-detuned
    edge of the optical resonance and normalised to unit peak height.  The dashed
    grey curves are Lorentzian fits.  The progressive narrowing of the mechanical
    linewidth with increasing power is a direct signature of optomechanical
    backaction amplification approaching the phonon-lasing threshold.
    (\textbf{d})~Effective mechanical linewidth $\Gamma_\mathrm{eff}/2\pi$ extracted
    from Lorentzian fits to the mechanical spectra as a function of corrected
    intracavity power $P_\mathrm{in}$ (see text).  The solid line is a linear fit;
    the dashed extension indicates the extrapolation to the phonon-lasing threshold
    $P_\mathrm{th} = 1.13$~mW (in cavity coordinates) where
    $\Gamma_\mathrm{eff} \to 0$.  The slope of the linear fit yields the
    single-photon optomechanical coupling rate $g_0/2\pi = 551.9 \pm 19.1$~kHz.
    Data points are colour-coded consistently with panels~(a) and~(c).}
    \label{fig:S_charact}
\end{figure}

Figure~\ref{fig:S_charact} summarises the optomechanical characterization of the
device using two complementary measurements.

Panel~(a) shows the normalised optical transmission spectra for ten input powers
ranging from 1.05 to 4.55~mW.  As the power increases, the resonance dip
red-shifts progressively due to thermo-optic heating of the silicon slab,
consistent with the coupled-mode model (Eq.~\ref{eq:S_thermal}).  The dashed line
marks the threshold transmission $\mathcal{T}_\mathrm{th} \approx 0.44$, which
corresponds to a laser--cavity detuning $\Delta \approx -\kappa/2$ and defines the
operating point used for all reservoir computing measurements.

Panel~(b) shows the optical transmission at very low power (0.04~mW), where
thermo-optic effects are negligible, together with a Lorentzian fit from which the
optical decay rates are extracted: $\kappa_\mathrm{total}/2\pi = 7.65$~GHz,
$\kappa_i/2\pi = 6.49$~GHz, and $\kappa_e/2\pi = 1.155$~GHz, confirming operation
in the undercoupling regime ($\kappa_e \ll \kappa_i$), with loaded quality factor
$Q_L = 25{,}174$.  These values enter the corrected intracavity power
\begin{equation}\label{eq:S_Pcav}
    P_\mathrm{cav} = P_\mathrm{in} \cdot \eta_\mathrm{EOM} \cdot
    \frac{4\kappa_e}{\kappa} \cdot T_\mathrm{loop},
\end{equation}
where $\eta_\mathrm{EOM}$ is the EOM power transmission,
$4\kappa_e/\kappa$ is the on-resonance power-coupling factor into
the cavity, and $T_\mathrm{loop} \approx 0.80$ accounts for tapered-fibre loop losses.

Panel~(c) shows normalised mechanical PSD spectra at three representative input
powers (2.55, 3.25, and 4.05~mW), colour-coded consistently with panel~(a).  The
progressive narrowing of the Lorentzian lineshape confirms increasing backaction
amplification as the phonon-lasing threshold is approached from below.

To extract $g_0$, we record the mechanical PSD at each power with the laser tuned
to the blue-detuned edge ($\Delta \approx -\kappa/2$), where optomechanical
backaction is dominantly amplifying and the effective mechanical damping rate takes
the form
\begin{equation}\label{eq:S_Gamma_eff}
    \Gamma_\mathrm{eff} = \Gamma_m + g_0^2 |\bar{a}|^2\,\mathcal{A}(\Delta),
\end{equation}
where $\Gamma_m$ is the intrinsic mechanical damping rate,
$|\bar{a}|^2 \propto P_\mathrm{cav}$ is the mean intracavity photon number, and
$\mathcal{A}(\Delta)$ is the backaction factor at the operating detuning.
Panel~(d) shows $\Gamma_\mathrm{eff}/2\pi$ as a function of $P_\mathrm{cav}$ for
all ten powers; a linear fit yields a phonon-lasing threshold
$P_\mathrm{th} = 1.13$~mW (in cavity coordinates) and a single-photon
optomechanical coupling rate $g_0/2\pi = 551.9 \pm 19.1$~kHz.

\begin{figure}[htbp]
    \centering
    \includegraphics[width=0.5\textwidth]{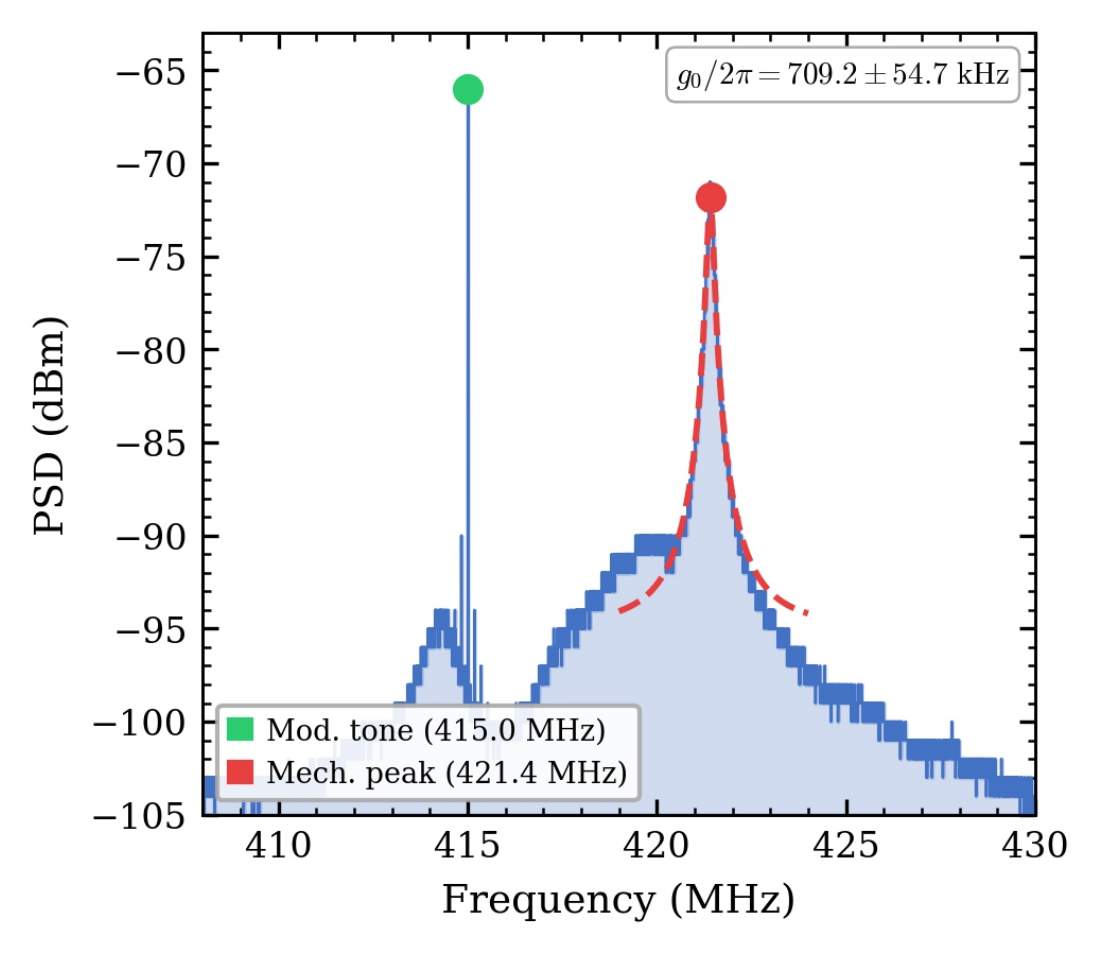}
    \caption{Phase-modulation calibration of the vacuum optomechanical coupling
    rate.  The RF power spectral density recorded in the sub-threshold (Brownian
    motion) regime shows the thermomechanical noise peak at
    $\Omega_m/2\pi \approx 421.4$~MHz (red dot) and a phase-modulation reference
    tone at $\Omega_\mathrm{mod}/2\pi \approx 415.0$~MHz (green dot) injected via
    the electro-optic modulator with known modulation depth $\phi$.  The spectrum
    is the average of 15 independent measurements.  The dashed red
    curve is a Lorentzian fit to the mechanical peak.  Comparing the integrated
    areas of the two peaks via Eq.~\eqref{eq:S_g0_phasemod} yields
    $g_0/2\pi = 709.2 \pm 54.7$~kHz.}
    \label{fig:S_phasemod}
\end{figure}

As an independent cross-check, $g_0$ is extracted by comparing the integrated
thermomechanical noise peak to a phase-modulation tone of known depth $\phi$
injected near $\Omega_m$ (Fig.~\ref{fig:S_phasemod}).  The spectrum is averaged
over 15 independent measurements to reduce the statistical uncertainty on the
peak areas.  Since the two frequencies
are sufficiently close, the transduction function of the detection chain is
approximately constant, and
\begin{equation}\label{eq:S_g0_phasemod}
    g_0 = \sqrt{\frac{S_\mathrm{mec}^{II}(\Omega_m)\,\Gamma_m/4}
                     {S_\mathrm{mod}^{II}(\Omega_\mathrm{mod})\,\mathrm{ENBW}}}
    \;\frac{\phi\,\Omega_\mathrm{mod}}{4\sqrt{n_\mathrm{th}}},
\end{equation}
where $S^{II}$ denotes the one-sided power spectral density, ENBW is the effective
noise bandwidth of the signal analyser, $n_\mathrm{th} = k_BT/(\hbar\Omega_m)$ is
the mean thermal phonon occupancy, and $\phi = \pi V_\mathrm{rf}/V_\pi$ is the EOM
modulation depth.  This method yields $g_0/2\pi = 709.2 \pm 54.7$~kHz, consistent
with the backaction estimate within the combined uncertainties arising primarily
from the tapered-fibre loop transmission $T_\mathrm{loop}$.

\subsection{Dynamical regimes of the optomechanical oscillator}

\begin{figure}[htbp]
    \centering
    \includegraphics[width=0.85\textwidth]{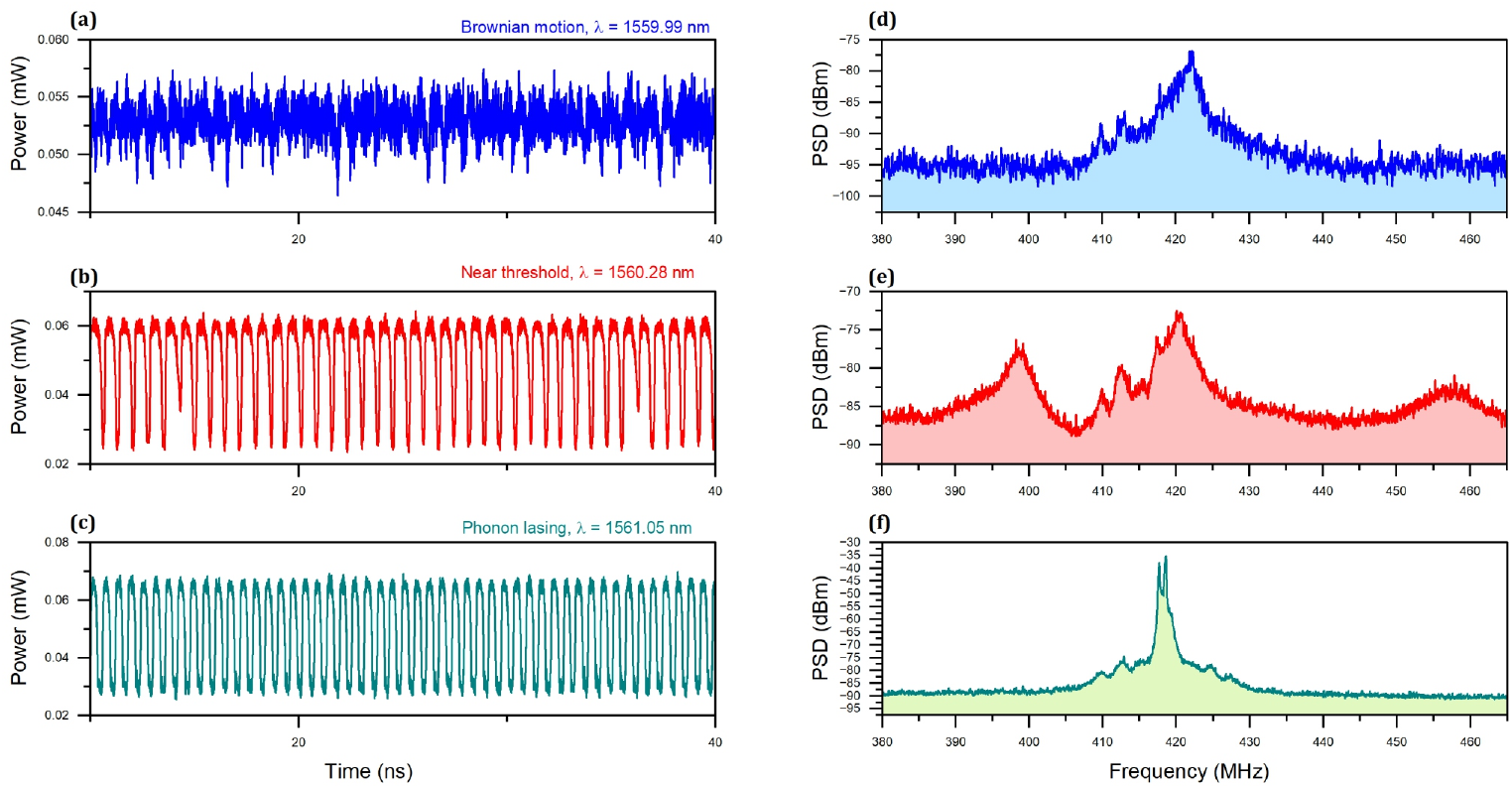}
    \caption{Dynamical regimes of the optomechanical cavity as a function of
    laser wavelength. Left column: time traces of the transmitted optical
    power recorded by the oscilloscope (sampling rate 20~GSa/s, decimated for
    display). Right column: corresponding RF power spectral density (PSD)
    measured by the signal analyser.
    (\textbf{a},\,\textbf{d})~Brownian motion regime
    ($\lambda_\mathrm{L} = 1559.99$~nm): the transmitted power fluctuates
    stochastically and the PSD shows a broad thermomechanical Lorentzian with
    no coherent peak.
    (\textbf{b},\,\textbf{e})~Near-threshold regime
    ($\lambda_\mathrm{L} = 1560.28$~nm): large irregular oscillations appear
    in the time domain and the PSD develops multiple broad spectral features,
    signaling the onset of nonlinear dynamics near the Hopf bifurcation.
    (\textbf{c},\,\textbf{f})~Phonon-lasing regime
    ($\lambda_\mathrm{L} = 1561.05$~nm): the cavity settles into a
    self-sustained limit cycle and the PSD collapses to a single narrow
    coherent tone more than 50~dB above the noise floor.}
    \label{fig:S_bifurcation}
\end{figure}

Figure~\ref{fig:S_bifurcation} illustrates the three distinct dynamical regimes
accessed by tuning the laser wavelength across the phonon-lasing threshold.
At $\lambda_\mathrm{L} = 1559.99$~nm, far from the blue-detuned edge of the
optical resonance, the intracavity photon number is insufficient to overcome
the intrinsic mechanical damping. The mechanical mode remains in its thermally
driven regime: the transmitted power fluctuates stochastically around its mean
value (Fig.~\ref{fig:S_bifurcation}a) and the corresponding RF power spectral
density exhibits a broad Lorentzian lineshape centred near
$\Omega_m/2\pi \approx 422$~MHz with no coherent feature
(Fig.~\ref{fig:S_bifurcation}d), consistent with Brownian motion of the
mechanical oscillator.

As the laser is tuned to $\lambda_\mathrm{L} = 1560.28$~nm, approaching the
phonon-lasing threshold from below, the optomechanical backaction amplification
partially compensates the mechanical damping. The transmitted power develops
large-amplitude, irregular oscillations (Fig.~\ref{fig:S_bifurcation}b) and
the RF spectrum broadens into multiple unresolved spectral features
(Fig.~\ref{fig:S_bifurcation}e), reflecting the complex nonlinear transient
dynamics characteristic of the pre-lasing regime near the Hopf bifurcation.

At $\lambda_\mathrm{L} = 1561.05$~nm, beyond the threshold $P_\mathrm{th} =
1.13$~mW (in cavity coordinates), the cavity enters the phonon-lasing regime. The transmitted power
exhibits regular, large-amplitude oscillations at a period set by the interplay
of the thermal and free-carrier nonlinearities (Fig.~\ref{fig:S_bifurcation}c),
and the RF spectrum collapses to a single sharp coherent tone at
$\Omega_m/2\pi = 418.6$~MHz rising more than 50~dB above the noise floor
(Fig.~\ref{fig:S_bifurcation}f). This self-sustained limit cycle constitutes
the phonon-lasing attractor that the reservoir computing scheme exploits.

\section{Experimental optimization of reservoir computing}
\subsection{Validation of the numerical model}

\begin{figure}[!htbp]
    \centering
    \includegraphics[width=0.8\textwidth]{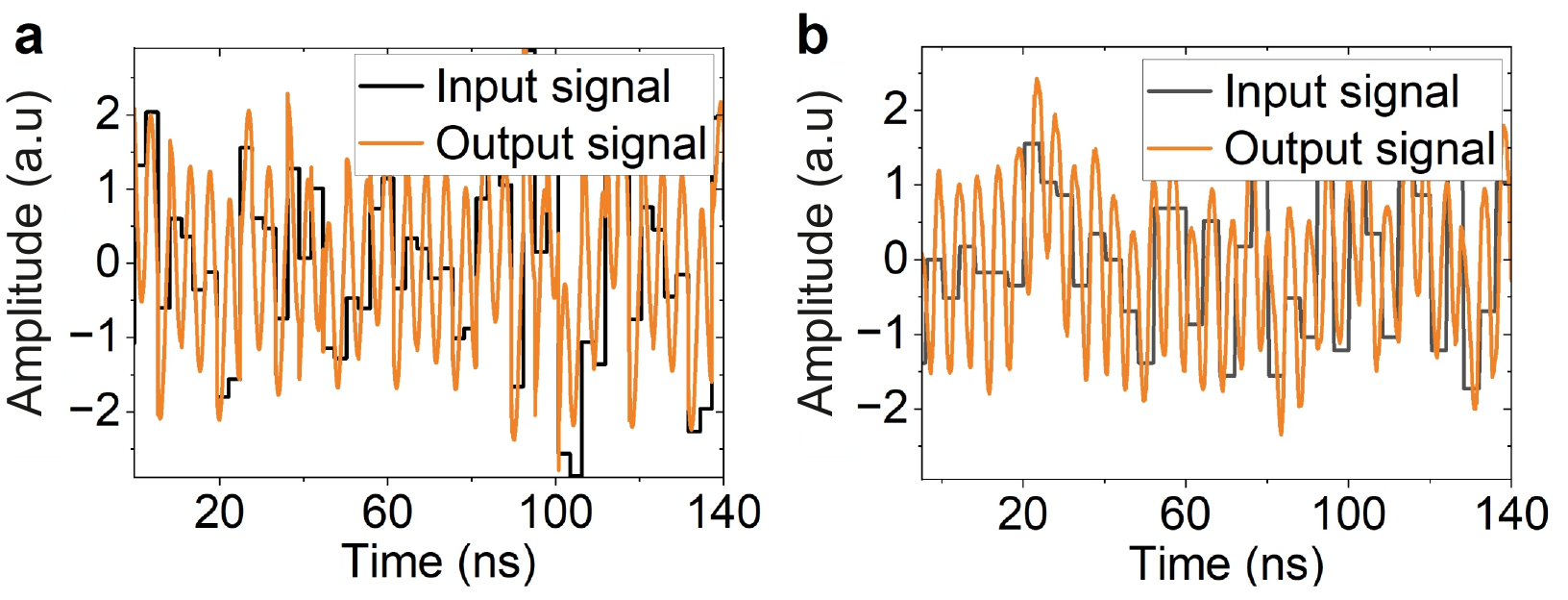}
    \caption{Input and output signals in the time-multiplexed reservoir for (a)~numerical simulation and (b)~experimental measurement. The black (grey) traces show the masked input signal, and the orange traces show the cavity output. The qualitative agreement between simulation and experiment confirms that the coupled-mode model (Eqs.~\ref{eq:S_mech}--\ref{eq:S_thermal}) captures the essential nonlinear dynamics of the optomechanical reservoir.}
    \label{fig:S_simvsexp}
\end{figure}

Figure~\ref{fig:S_simvsexp} compares the reservoir input and output signals obtained from the numerical simulation and the experiment. The simulation parameters are chosen to match the experimental conditions: the DC input power is $\bar{P}_\mathrm{in} = 0.98$~mW, the perturbative modulation power is $\delta P_\mathrm{in} = 0.02$~mW, and the detuning is $\delta\omega = 0$. These values ensure that the cavity operates in the regime of self-sustained optomechanical oscillations, where the reservoir input signal acts only as a small perturbation that does not disrupt the coherent mechanical motion.

In both cases, the output signal exhibits the fast oscillation at the mechanical frequency modulated by the slower envelope of the time-multiplexed input. The qualitative agreement between simulation and experiment confirms that the coupled-mode model (Eqs.~\ref{eq:S_mech}--\ref{eq:S_thermal}) captures the essential nonlinear dynamics exploited for reservoir computing.

\subsection{Optimization of modulation strength and hold time}

\begin{figure}[htbp]
    \centering
    \includegraphics[width=0.7\textwidth]{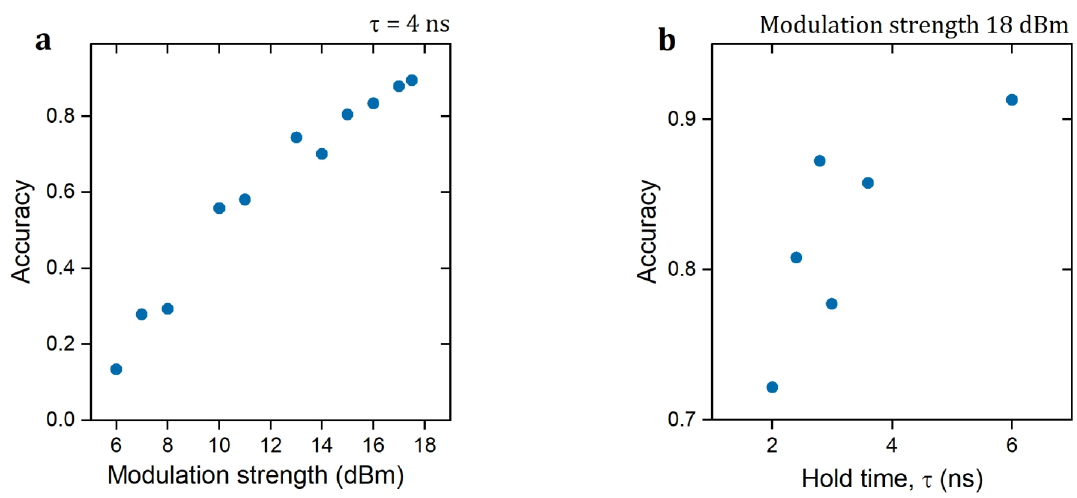}
    \caption{Optimization of the reservoir operating parameters. (\textbf{a})~Accuracy (Pearson correlation between the cavity transmitted signal and the modulating signal) as a function of the modulation strength. The accuracy increases monotonically with increasing modulation strength, indicating that the reservoir performs best when the input signal is as high as possible while still acting as a weak perturbation of the phonon-lasing attractor. (\textbf{b})~Accuracy as a function of the hold time $\tau$. The performance is broadly optimal for hold times in the range $\tau \approx 2$--$6$~ns.}
    \label{fig:S_optimization}
\end{figure}

Two key parameters govern the performance of the time-multiplexed reservoir: the modulation strength of the input signal and the hold time $\tau$ assigned to each virtual node. We optimize both experimentally by measuring the accuracy, defined as the Pearson correlation between the input data and the signal transmitted  by the cavity. Where the input data is a set of random numbers held constant for a period $\tau$, and the signal transmitted by the cavity is averaged over period $\tau$ (see Fig. \ref{fig:S_readout}).
Figure~\ref{fig:S_optimization}a shows the accuracy as a function of the modulation strength. The highest accuracy ($\sim$0.9) is achieved at the highest modulation strength, taking care that the input signal is still a weak perturbation that probes the nonlinear dynamics of the phonon-lasing attractor without disrupting it. As the modulation strength decreases, the accuracy degrades progressively: below $\sim$11~dBm the accuracy drops below 0.6, and at 6~dBm it falls to $\sim$0.13. This behavior is consistent with the perturbative operating principle of the reservoir---the coherent limit cycle must remain the dominant dynamical feature for the cavity to perform useful nonlinear transformations.

Figure~\ref{fig:S_optimization}b shows the optimal window for accuracy as a function of the hold time $\tau$. Accuracy above 0.8 is reached at $\tau=2.6$ ns, see next section for further discussion.

\subsection{Effect of node duration time on reservoir dynamics}

\begin{figure}[htbp]
    \centering
    \includegraphics[width=0.95\textwidth]{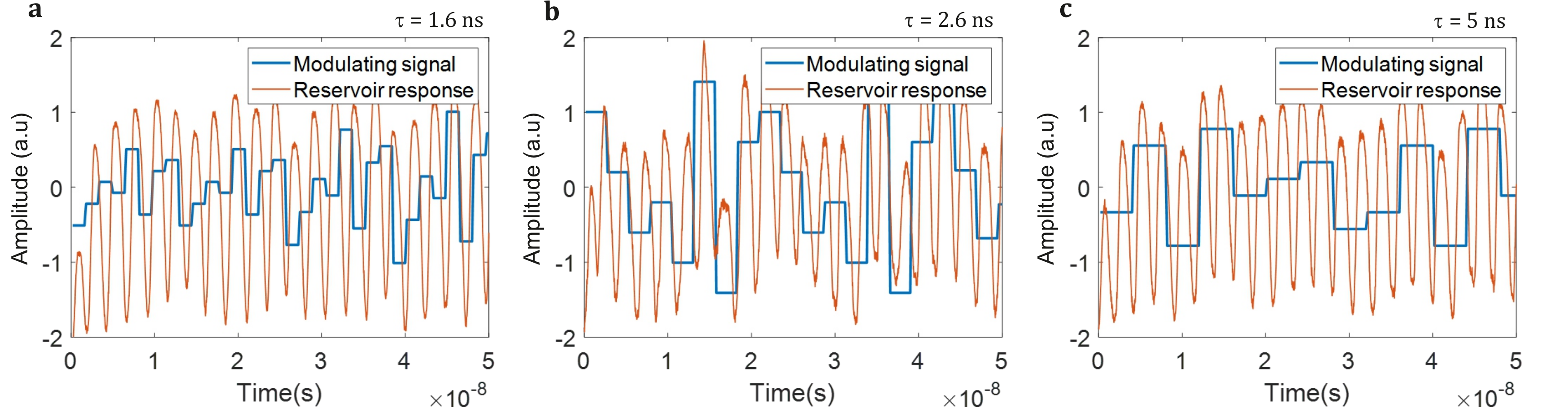}
    \caption{Modulating signal (blue) and reservoir response (orange) for three node separation times: (\textbf{a})~$\tau = 1.6$~ns, (\textbf{b})~$\tau = 2.6$~ns, and (\textbf{c})~$\tau = 5$~ns. At 1.6~ns the node separation is shorter than the mechanical period $T_m = 2.39$~ns and the reservoir cannot complete a full oscillation cycle between consecutive nodes, undersampling the dynamics. At 2.6~ns the node separation matches $T_m$, and the reservoir response resolves each virtual node with maximal nonlinear contrast. At 5~ns the node separation exceeds $T_m$ and approaches the thermal dissipation lifetime, so that thermal relaxation begins to smooth the response between consecutive nodes.}
    \label{fig:S_timetraces}
\end{figure}

The physical origin of the hold-time/node duration time  optimum is revealed in Fig.~\ref{fig:S_timetraces}, which shows the modulating signal and the reservoir response for three representative hold times. At $\tau = 1.6$~ns (Fig.~\ref{fig:S_timetraces}a), the hold time is shorter than the mechanical period $T_m = 2.39$~ns: the reservoir cannot complete a full oscillation cycle between consecutive nodes, and the response undersamples the dynamics. At $\tau = 2.6$~ns (Fig.~\ref{fig:S_timetraces}b), the node separation matches $T_m$, and each virtual node elicits a distinct nonlinear transient that fills the mechanical phase space---this is the regime of maximal information encoding. At $\tau = 5$~ns (Fig.~\ref{fig:S_timetraces}c), the node separation exceeds $T_m$ and approaches the thermal dissipation lifetime ($\tau_\mathrm{th} \sim 10$~ns), so that thermal relaxation begins to smooth the response and erase the contrast between consecutive nodes. Based on this optimization, all results reported in the main text use a modulation strength of 18-19 dBm and node duration time in the optimal regime of 2-4 ns.

\subsection{Post-processing}

\begin{figure}[htbp]
    \centering
    \includegraphics[width=0.8\textwidth]{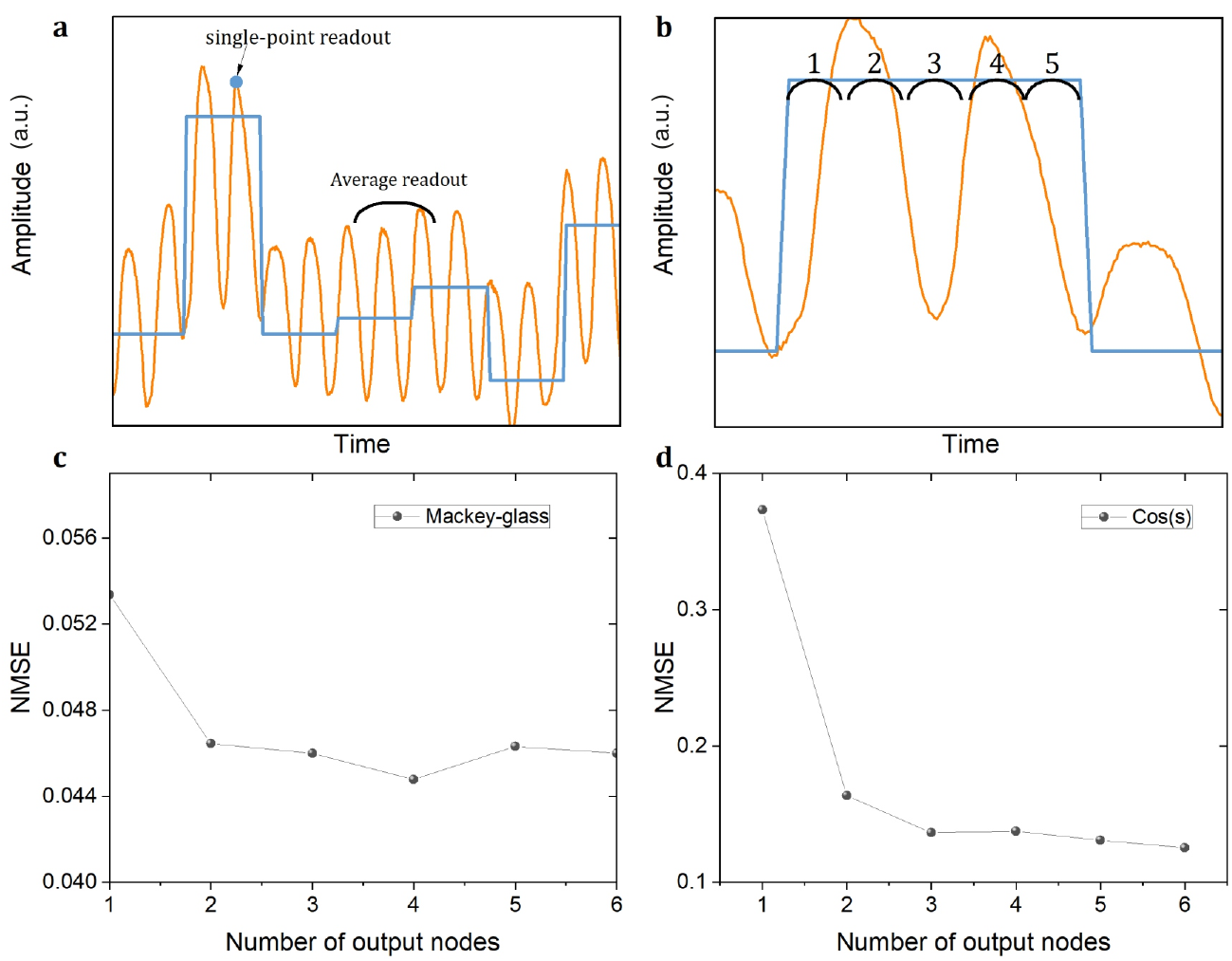}
    \caption{(a). Schematic of the readout strategy. Because the reservoir output oscillates at the mechanical frequency within each node interval, a single-point readout (dot) is sensitive to the sampling phase. Instead, each virtual node state is extracted by averaging the transmitted signal over the full node duration (bracket), which integrates over the fast mechanical oscillation and retains only the amplitude envelope shaped by the input. (b) Division of readout signal into additional output nodes, increasing memory and dimensionality of the reservoir. NMSE as a function of the number of output nodes, number of input nodes is 20 for (c) Mackey-glass 1-step prediction task and (d) cosine function reconstruction. }
    \label{fig:S_readout}
\end{figure}

The oscillatory nature of the reservoir response also dictates the readout strategy. Unlike delay-line reservoirs where the output is slowly varying and can be sampled at a single point per virtual node, the output of the optomechanical reservoir oscillates at the mechanical frequency within each node interval (Fig.~\ref{fig:S_readout}). A single-point readout is therefore sensitive to the sampling phase and yields degraded performance. We instead extract each virtual node state by averaging the transmitted signal over the full node duration $\tau$, which integrates over the fast mechanical oscillation and retains only the amplitude envelope shaped by the input.

The sampling rate of the oscilloscope was set to 10GSa/s. This means that for acquisition time of 4$\mu$s (which is the coherence time of the cavity), a node with a duration of 2.4 ns, has 6,000 corresponding samples in the read-out signal. This enables us to extend the output layer, adding nodes and increasing both the number of trained output weights and the effective memory. For instance, if the input mask has 20 nodes, we could take each readout sample as a node and extend the output layer to $20 \times 6000$ nodes.
We have found that the optimal number of output nodes is 2-6, where we divide the read-out samples into 2-6 equal parts and average over each separately.
In Fig. \ref{fig:S_readout}(c) and (d) we plot the NMSE for the Mackey-Glass 1-step prediction task and for cosine function reconstruction vs. number of output nodes. For both tasks, optimum is reached around 4 output nodes.

\subsection{Nonlinear function reconstruction}\label{sec:NL}
\begin{figure}[htbp]
    \centering
    \includegraphics[width=0.6\textwidth]{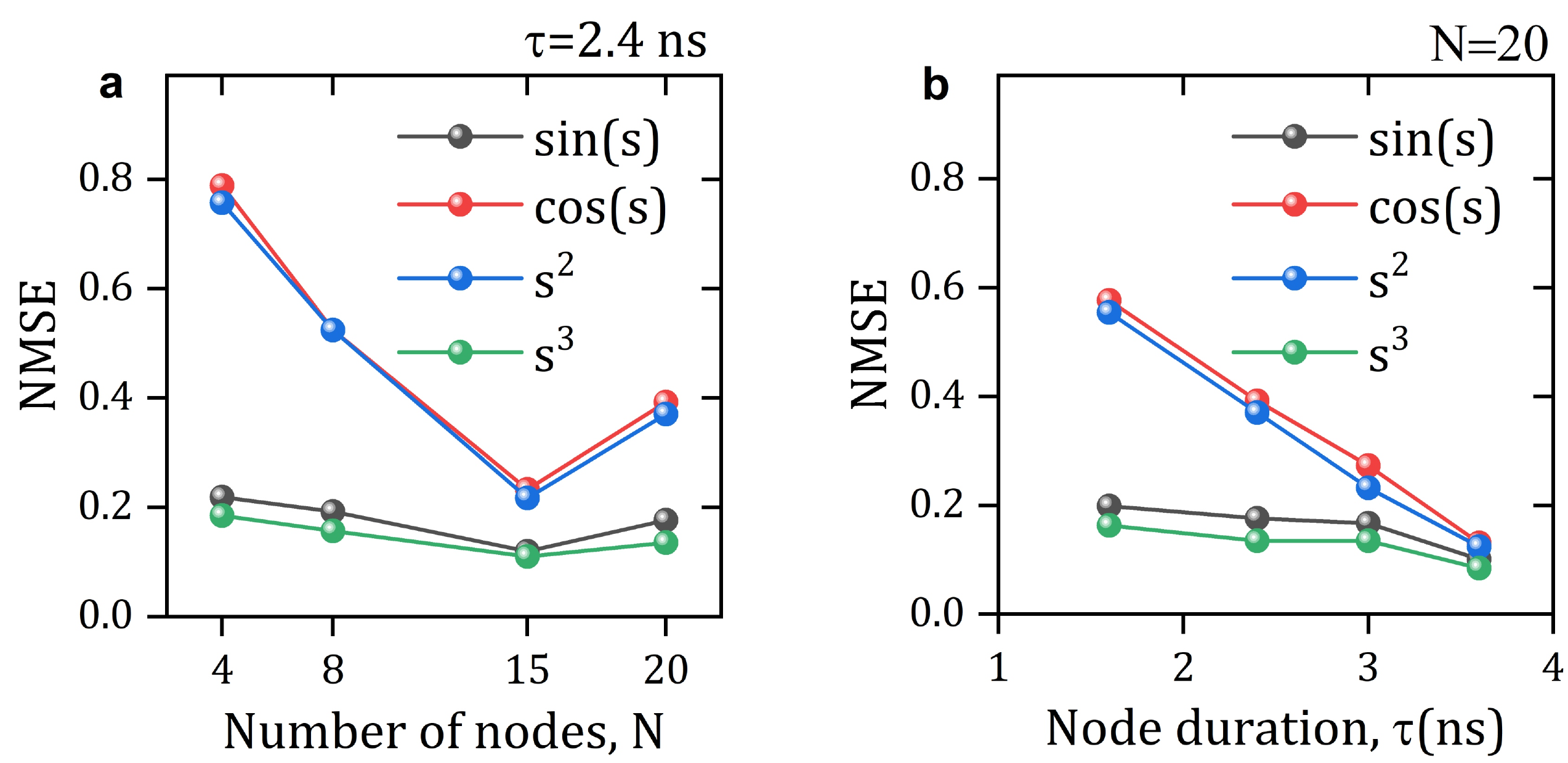}
    \caption{Nonlinear function reconstruction performance.
    (\textbf{a})~NMSE as a function of the number of virtual nodes $N$
    at fixed $\tau = 2.4$~ns. (\textbf{b})~NMSE as a function of hold
    time $\tau$ at fixed $N = 20$. In both panels, odd target functions
    ($\sin(s)$, $s^3$) show systematically lower NMSE than even functions
    ($\cos(s)$, $s^2$); the physical origin of this asymmetry under power
    modulation is discussed in Section~\ref{sec:NL}.}
    \label{fig:S_NL}
\end{figure}

Figure~\ref{fig:S_NL} quantifies the reconstruction accuracy of four
nonlinear target functions (see Fig.~\ref{fig:2}) as a function of the
number of virtual nodes $N$ (at fixed $\tau = 2.4$~ns) and as a function
of the hold time $\tau$ (at fixed $N = 20$). The NMSE is systematically
lower for odd functions across all operating conditions: at $N = 20$ and
$\tau = 3.6$~ns, we obtain NMSE~$\approx 0.1$ for $\sin(s)$ and
NMSE~$\approx 0.08$ for $s^3$, compared with NMSE~$\approx 0.13$ for
$\cos(s)$ and NMSE~$\approx 0.125$ for $s^2$. Both function classes show
improved performance with increasing $\tau$, consistent with the deeper
memory integration discussed in Section~\ref{sec:memory}. The reconstruction
accuracy and the odd/even asymmetry together characterize the nonlinear
response of the cavity around the phonon-lasing operating point as
predominantly low-order, consistent with weak perturbations of the coherent
attractor. In systems where the input is encoded as a laser frequency
detuning, the odd/even asymmetry has a natural geometric explanation: the
Lorentzian cavity transmission is antisymmetric around the operating point
at $\Delta \approx -\kappa/2$, so detuning modulation maps directly onto an
antisymmetric transduction function, favouring odd targets. In the present
experiment, however, the input is encoded as a power modulation. The response
symmetry is then determined by the full nonlinear dynamics of the coupled
system --- radiation pressure, free-carrier dispersion, and thermo-optic
effects --- rather than by the slope of the Lorentzian alone. How the
effective symmetry of the transduction depends on the operating point,
modulation depth, and the interplay between these nonlinear mechanisms under
power modulation remains an open question that will be the subject of future
work.

\section{Numerical simulation of memory capacity}\label{sec:memory}

\begin{figure}[t!]
    \centering
    \includegraphics[width=1\textwidth]{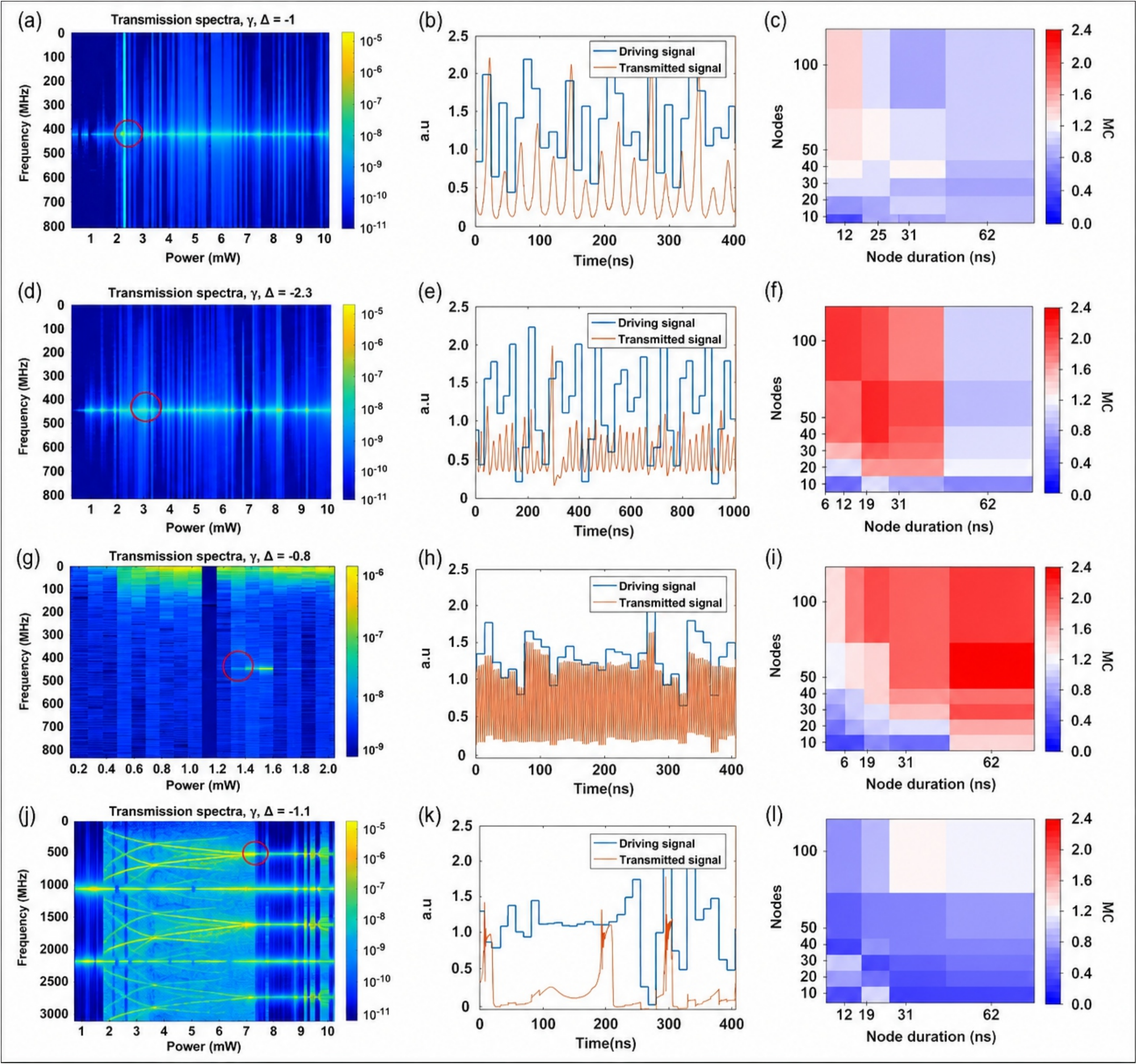}
    \caption{Memory capacity of the optomechanical cavity reservoir from numerical simulations. (a) and (d)~Transmission spectra of the optomechanical cavity with parameters used in this experiment ($\Omega_m/2\pi = 418$~MHz, $g_0/2\pi = 770$~kHz, $\gamma_i/2\pi = 4$~GHz, $\gamma_e/2\pi = 1.155$~GHz) with detunings $\Delta/\gamma_i = -1$ and $\Delta/\gamma_i = -2.3$ respectively. (c) and (f)~Average memory capacity for the dynamical regimes of self-sustained oscillations (red circled in (a) and (d)). (g)~Transmission spectra of the same optomechanical cavity without the thermo-optic effect. (i)~Average reservoir memory capacity for the cavity without the thermo-optic effect; the memory capacity is comparable to that obtained when including the thermo-optic effect. (j)~Transmission spectra with sub-harmonics of an optomechanical cavity with mechanical period close to the thermal dissipation lifetime ($\Omega_m/2\pi = 110$~MHz, $g_0/2\pi = 690$~kHz, $\gamma_i/2\pi = 19$~GHz, $\gamma_e/2\pi = 2.2$~GHz) and detuning $\Delta/\gamma_i = -1.1$. (l)~Average reservoir memory capacity for the cavity with parameters in (j) and an input power of 8~mW, selected so that the cavity is in self-sustained oscillations with the second harmonic. (b, e, h, k)~Signal transmitted from the cavity and modulating signal. In (b) and (k), the cavity adheres to its natural oscillations more strongly than to the external modulating signal, resulting in lower memory capacity than in (e) and (h), where the cavity follows the modulating signal.}
    \label{fig:S_MC}
\end{figure}

The coupled equations~\eqref{eq:S_mech}--\eqref{eq:S_thermal} are solved numerically using a fourth-order Runge-Kutta algorithm after normalizing time by $\kappa = \gamma_i + \gamma_e$, i.e.\ $\hat{t} = \kappa t / 2$ and $\hat{x} = G x / \kappa$, where $G = g_0 \sqrt{2 m_\mathrm{eff} \Omega_m / \hbar}$. A time step of $0.1$ was used for the integration.

The linear memory capacity (MC) quantifies the ability of the reservoir to recall past input values from its current state. It is defined as:
\begin{equation}\label{eq:S_MC}
    \mathrm{MC} = \sum_{p=1}^{P} \mathrm{corr}\!\left(s_p(k),\, \hat{o}_p(k)\right),
\end{equation}
where $\hat{o}_p(k)$ is the value recalled by the reservoir for $p$ previous time steps, $s_p(k)$ is the actual past input value, and $\mathrm{corr}(\cdot, \cdot)$ denotes the Pearson correlation. The input data $s(k)$ is a sequence of random numbers drawn from a uniform distribution, with 2{,}000 data points for training and 1{,}000 for testing. We compute Eq.~\eqref{eq:S_MC} up to $P = 5$.

The memory observed here arises from the coexistence of multiple dynamical timescales in the system and from the inertia of the mechanical oscillations. The fast mechanical motion, together with slower carrier and thermal dynamics, enables the encoding of a temporally weighted history of past inputs. According to this model, the maximum attainable memory capacity could reach about two-to-three past steps when combining the right dynamical regime and the mechanical period.

In Figure~\ref{fig:S_MC}a and d we show the transmission spectra of a cavity with the same parameters as were used in the experiment ($\Omega_m/2\pi = 418$~MHz, $g_0/2\pi = 770$~kHz, $\gamma_i/2\pi = 4$~GHz, and $\gamma_e/2\pi = 1.15$~GHz) and with detunings $\Delta/\gamma_i = -1$ and $\Delta/\gamma_i = -2.3$ respectively. The red circle encapsulates a dynamical regime of self-sustained oscillations (phonon lasing) with very high amplitude (a) and slightly lower amplitude (d). Figures~\ref{fig:S_MC}c and f show the average memory capacity (each value is an average over five simulations with different random masks) for various values of node duration and number of nodes, when the reservoir operates in the self-sustained oscillation regimes described in (a) and (d). In these dynamical regimes the maximum memory capacity could reach 2--3 past steps.

When removing the thermo-optic effect and considering only the optomechanical coupling, the model reduces to two equations: Eq.~\eqref{eq:S_mech} is unchanged, while Eq.~\eqref{eq:S_optical} simplifies to:
\begin{equation}\label{eq:S_optical_notherm}
\begin{aligned}
    \frac{dU}{dt} &= i\biggl(-g_0 \sqrt{\frac{2 m_\mathrm{eff} \Omega_m}{\hbar}} x(t) +  \delta\omega \biggr) U(t) &- \frac{1}{2}\left(\gamma_i + \gamma_e\right) U(t) &+ \sqrt{\gamma_e \left(\bar{P}_\mathrm{in} + \delta P_\mathrm{in}\, U_\mathrm{inj}(t)\right)}.
\end{aligned}
\end{equation}
We solved this model with the experimental cavity parameters. The transmission spectra show self-sustained oscillations with detuning $\Delta/\gamma_i = -0.8$ and input power of 1.3~mW, see Figure~\ref{fig:S_MC}g. The memory capacity reaches values similar to those obtained when the thermo-optic effect is included, see Figure~\ref{fig:S_MC}i.

To further illustrate the richness of the system, we simulated an optomechanical cavity with different parameters ($\Omega_m/2\pi = 110$~MHz, $g_0/2\pi = 690$~kHz, $\gamma_i/2\pi = 19$~GHz, and $\gamma_e/2\pi = 2.2$~GHz) where the mechanical period ($T_m \approx 9$~ns) approaches the thermal dissipation lifetime. This type of cavity enables richer dynamics with self-sustained oscillations that exhibit sub-harmonics, see Figure~\ref{fig:S_MC}j. When operating in the second-harmonic dynamical regime, the cavity strongly adheres to its natural oscillations and nearly ignores the external modulation of the reservoir (Figure~\ref{fig:S_MC}k); the average memory capacity is therefore limited to about 1, see Figure~\ref{fig:S_MC}l.

Figures~\ref{fig:S_MC}c, f, i, and l reveal that, for a given mechanical frequency and dynamical regime, the memory capacity depends on the node duration $\tau$: longer node durations allow the reservoir to integrate information over more mechanical oscillation cycles within each input step, systematically increasing memory. However, there is an optimum value beyond which memory capacity starts to decrease. Similarly, increasing the number of nodes increases the memory capacity, as the maximum memory capacity is theoretically bounded by the number of nodes in the reservoir~\cite{Jaeger2002_SI}.

\section{Benchmarking}
\subsection{Prediction of Mackey-Glass chaotic time series}
The Mackey-Glass time series is obtained by integrating in time the following equation:
\begin{equation}
    \frac{dx(t)}{dt}=\frac{\alpha x(t-\tau)}{1+x(t-\tau)^\beta}-\gamma x(t)
    \label{eq:mg}
\end{equation}
and the following parameter values: $\alpha=0.2$, $\beta=10$, $\gamma=0.1$, and $\tau=17$.
The integration step is of 0.1 to which we apply an oversampling of a factor of 3.
The solution of Eq. \eqref{eq:mg} describes a weakly chaotic behavior.
The prediction of this Mackey-glass time series is a difficult computational task, as it requires both nonlinear transformation and memory capacity.\\

\subsection{Spoken digit recognition}

Spoken digit recognition tests the reservoir's capacity for nonlinear classification over structured, real-world input data with inherent temporal correlations and spectral complexity, providing a complementary benchmark to the Mackey--Glass time-series prediction task. We use the TI-46 speech corpus~\cite{TI46_SI}, a standard benchmark for reservoir computing evaluation~\cite{Verstraeten2005_SI}, consisting of isolated recordings of the digits 0--9 spoken by five speakers (10 repetitions each, 500 utterances total). Each audio waveform is preprocessed with the Lyon cochlear model~\cite{Lyon1982}, which converts the raw waveform into a cochleagram: an $86 \times K$ frequency--time matrix representing the spectral power distribution over 86 channels, where $K$ ranges from 32 to 130 time steps depending on the utterance duration.

To inject this two-dimensional input into the single-node time-multiplexed reservoir, each cochleagram is multiplied by a random mask of dimension $N \times 86$, projecting the 86-dimensional frequency space onto the $N$-dimensional virtual-node space. The resulting $K \times N$ matrix is serialized into a one-dimensional time-multiplexed signal using the same hold-time protocol as the Mackey--Glass task: each masked value is held for a duration $\tau$. The reservoir operates in the same perturbative phonon-lasing regime as in the other tasks, where the input signal acts as a weak modulation of the underlying coherent attractor without disrupting the limit cycle. The linear readout is trained by ridge regression against a $10 \times K$ target matrix in which element $(c,k)$ is $+1$ if digit $c$ was spoken at step $k$ and $-1$ otherwise. Classification selects the class with the highest time-averaged readout score, and performance is reported as the word error rate (WER).

The maximum coherent oscillation time of the phonon-lasing state is 4~$\mu$s (Section~IV). For $N = 19$ virtual nodes and hold time $\tau = 2.4$~ns, each input step occupies $N\tau = 45.6$~ns, so an average $K = 80$-step utterance requires $\sim$3.6~$\mu$s of injection time — as such we need to run a whole measurement for each utterance, i.e. 500 measurements. In order to make this task less daunting, we therefore evaluate performance on a two-digit classification sub-task (for digits 1 and 2, 100 utterances in total). Extending the coherent oscillation time through higher-$Q$ cavities or active phase stabilization would enable the full ten-class task.

\subsection{Energy per operation}

The energy consumed per virtual-node operation is estimated from the total optical power and the hold time per node. The cavity is driven at $P_\mathrm{in} = \bar{P}_\mathrm{in} + \delta P_\mathrm{in} \approx 1$~mW, where $\bar{P}_\mathrm{in} = 0.98$~mW sustains the self-oscillation and $\delta P_\mathrm{in} = 0.02$~mW carries the input signal. At the optimal hold time $\tau = 3$~ns identified in Fig.~\ref{fig:4}a:
\begin{equation}\label{eq:S_energy}
    E_\mathrm{node} = P_\mathrm{in} \times \tau = 1~\mathrm{mW} \times 3~\mathrm{ns} = 3~\mathrm{pJ}.
\end{equation}
This accounts for the full optical energy delivered to the device, including the power required to sustain the coherent nonlinear state that underlies the computation. Considering only the signal power ($\delta P_\mathrm{in} = 0.02$~mW) gives $\sim$60~fJ per operation, but the drive power cannot be excluded from a fair accounting since it is essential to maintain the attractor. In this sense, the same physical process—the coherent optomechanical oscillation—simultaneously provides the nonlinear transformation and the dynamical memory, so that no additional energy is required for separate feedback or storage mechanisms. Unlike delay-line reservoirs, where memory is implemented through a delayed feedback loop that may require additional resources to compensate losses, the entire energy budget here sustains the oscillation, performs the nonlinear transformation, and establishes the dynamical memory. This estimate refers to the optical power delivered to the device and does not include external amplification or detection electronics.

\subsection{Comparison with other photonic reservoir implementations}

Table~\ref{tab:comparison} compares the Mackey--Glass prediction performance of our implementation against other experimental photonic reservoirs. In all reference works, memory is engineered through an optical feedback loop and in some cases augmented by post-processing. Our system — with only 20 virtual nodes, no optical feedback, and $\sim$3~pJ per node operation — achieves competitive accuracy on both one- and two-step-ahead prediction.

\begin{table}[t!]
\centering
\begin{tabular}{|c|c|c|c|}
\hline
System & NMSE ($p=1$) & NMSE ($p=2$) & Time step \\
\hline
Ring resonator~\cite{Donati2024_TimeDelayReservoir} & 0.036 & 0.11 & 3 \\
\hline
Optoelectronic~\cite{picco2025efficient} & 0.063 & N/A & 1 \\
\hline
Semiconductor laser~\cite{bueno2017conditions} & 0.019 & 0.045 & 3 \\
\hline
Current work & 0.04 & 0.16 & 3 \\
\hline
\end{tabular}
\caption{Experimental RC implementations tested on the Mackey--Glass task.}
\label{tab:comparison}
\end{table}

\end{document}